\documentclass[]{aa}  

\usepackage{natbib}
\bibpunct{(}{)}{;}{a}{}{,}
\usepackage{graphicx}
\usepackage{revsymb}	
\usepackage{amsmath}	
\usepackage[usenames]{color}	
\usepackage{epstopdf}	
\usepackage{txfonts}
\usepackage{amssymb}
\usepackage{color}
\usepackage[justification=centering]{caption}
\usepackage{geometry} 
\usepackage[parfill]{parskip} 
\usepackage{amssymb}
\usepackage{slantsc}
\usepackage{array}
\usepackage{url}
\usepackage{amsfonts}
\usepackage[colorlinks=true,linkcolor=blue, urlcolor=blue, citecolor=blue]{hyperref}
\usepackage{sidecap}
\usepackage{graphicx}
\usepackage{xcolor}
\usepackage{nicefrac}
\usepackage{subfigure}
\usepackage{epsfig}
\colorlet{rouge}{red!70!darkgray}

\begin{document}
\title{MCMC inversions of the internal rotation of \textit{Kepler} subgiants}
\author{G. Buldgen\inst{1,2} \and L. Fellay\inst{2} \and J. Bétrisey\inst{1} \and S. Deheuvels\inst{3} \and M. Farnir \inst{2}  \and E. Farrell\inst{1}}
\institute{D\'epartement d'Astronomie, Universit\'e de Gen\`eve, Chemin Pegasi 51, CH-1290 Versoix, Switzerland \and STAR Institute, Universit\'e de Li\`ege, Li\`ege, Belgium \and IRAP, Universit\'e de Toulouse, CNRS, CNES, UPS, 14 avenue Edouard Belin, 31400, Toulouse, France }
\date{November, 2023}
\abstract{The measurement of the internal rotation of post-main sequence stars using data from space-based photometry missions has demonstrated the need for an efficient angular momentum transport in stellar interiors. So far, no clear solution has emerged and explaining the observed trends remain a challenge for stellar modellers.}
{We aim at constraining both the shape of the internal rotation profile of six \textit{Kepler} subgiants studied in details in 2014 and the properties of the missing angular momentum transport process acting in stellar interiors from Markov Chain Monte Carlo (MCMC) inversions of the internal rotation.}
{We apply a new MCMC inversion technique to existing \textit{Kepler} subgiant targets and test various shapes of the internal rotation profile of all six original subgiants observed in 2014. We also constrain the limitations on the number of free parameters that can be used in the MCMC inversion, showing the limitations in the amount of information in the seismic data.}
{First, we show that large-scale fossil magnetic fields are not able to explain the internal rotation of subgiants, similarly to what was determined from detailed studies of \textit{Kepler} red giants. We are also able to constrain the location of the transition in the internal rotation profile for the most evolved stars in the available set of subgiants. We find that some of them exhibit a transition located close to the border of the helium core while one clearly does not.}{We conclude that it might be possible that various processes might be at play to explain our observations, but that revealing the physical nature of the angular momentum process will require a consistent detailed modelling of all subgiants available, particularly the least evolved. In addition, increasing the number of stars for which such inferences are possible (e.g. with the future PLATO mission) is paramount given the key role they play in validating transport process candidates.}
\keywords{asteroseismology – stars: interiors – stars: evolution – stars: rotation}
\maketitle
\section{Introduction}

With the advent of space-based photometry missions such as CoRoT \citep{Auvergne2009}, \textit{Kepler} \citep{Borucki2010} and TESS \citep{Ricker2015}, stellar modellers gained direct access to constraints on the internal rotation of post-main sequence stars. Further data is expected to be made available with the future PLATO mission \citep{Rauer2014}. This was made possible by the observations of so-called mixed oscillation modes, which present a dual nature of both gravity and acoustic modes, thus bearing information on the innermost layers of these stars. Inference results \citep[e.g.][]{Beck2012,Deheuvels2012,Mosser2012,Deheuvels2014,Deheuvels2015,DiMauro2016,Gehan2018,DiMauro2018,Deheuvels2020,Fellay2021} have provided direct evidence of the presence of an efficient angular momentum transport process acting in stellar radiative zones. Indeed, stellar evolutionary models have proven unable to reproduce these constraints \citep{Marques2013} even if ``classical'' solutions put forward to reproduce the internal rotation of the Sun are introduced in the angular momentum transport equation \citep{Gough1998,spr02,Charbonnel2005}. Over the years, various candidates were put forward, none of them being able to provide a full solution \citep{Aerts2019}. 

An interesting feature of these various candidates lies in the shape of the internal rotation profile they would induce. For example, fossil magnetic fields as introduced by \citet{Kissin2015} and tested by \citet{Takahashi2021} would leave a solid body rotation in the radiative zone followed by a power law in radius in the outer convective envelope. Similarly, internal gravity waves and magnetic instabilities would be hindered by the presence of mean molecular weight gradients. Therefore, the transition from the slowly rotating envelope to the fast rotating core in their internal rotation profile should be located close to regions of strong chemical gradients \citep[see e.g.][for a classical textbook on the topic]{Maeder2009}. 

In subgiant and red giant stars, such regions are those characterized by the border of the helium core and the hydrogen shell. Therefore, some favoured shapes of rotation profiles can be tested and optimized to validate or invalidate the physical nature of the missing angular momentum transport mechanism. Such attempts were made for red giant branch stars in \citet{DiMauro2016}, \citet{DiMauro2018}, \citet{Fellay2021} and \citet{Wilson2023} on one of the subgiants studied here. While the former two privileged a transition located close to the hydrogen shell, the latter used an independent surface rotation by \citet{Garcia2014} to test a two-zone model and weakly constrain the location of the transition in rotation under $0.4$ normalized radii. MCMC inversions were also applied to KIC11145123 in \citet{Hatta2022}, which confirmed their SOLA inversions carried out in a previous publication \citep{Hatta2019}.

In this paper, we extend the experiment to the entire set of subgiant stars studied by \citet{Deheuvels2014}, using the original models and multiple shapes of reference rotation profiles. We start by briefly recalling the properties of the models in Sect. \ref{Sec:Models}, then carry out SOLA inversions in Sect \ref{Sec:SOLAInversions} and compare our results to \citet{Deheuvels2014}. In Sect \ref{Sec:MCMC}, we carry out an extensive analysis using MCMC inversions first on artificial data then on the observed splittings, discuss the properties of the profiles. In Sect. \ref{Sec:Disc}, we discuss our findings the implications for the angular momentum transport process acting on the subgiant branch and the limitations of our approach. We conclude and provide some additional perspectives in Sect. \ref{Sec:Conc}. 

\section{Stellar models}\label{Sec:Models}

The stellar models used in this study are the CESAM2K models of \citet{Deheuvels2014}, the oscillation frequencies and rotation kernels were computed using the ADIPLS oscillation package \citep{JCDAdipls2011}. We summarize their global properties in Table \ref{tabModels}. An interesting property of all the stars in this sample is that they have quite similar masses, allowing to analyse their rotation properties in an almost purely evolutionary context. However, we have to separate star F of the other stars as it is the only star which did not harbour a convective core during the main-sequence as well as star E which is at the other "extreme" of the considered mass range.

\begin{table*}[h]
\caption{Global parameters of the CESAM2K stellar evolutionary models from \citet{Deheuvels2014}}
\label{tabModels}
  \centering
\begin{tabular}{r | c | c | c | c | c | c }
\hline \hline
\textbf{Name}& KIC &$M/M_{\odot}$&Age (Gyr)&$R/R_{\odot}$& $(Z/X)_{0}$&Y$_{0}$\\ \hline
Star A& KIC 12508433 & $1.22$&$5.9$& $2.231$ & $0.0500$&$0.30$\\
Star B& KIC 8702606 &$1.27$&$3.8$& $2.467$ & $0.0173$&$0.27$\\ 
Star C& KIC 5689820 &$1.14$&$6.9$& $2.297$ & $0.0388$&$0.30$\\
Star D& KIC 8751420 &$1.26$&$3.8$& $2.670$ & $0.0151$&$0.27$\\
Star E& KIC 7799349 &$1.39$&$3.8$& $2.829$ & $0.0548$&$0.30$\\
Star F& KIC 9574283 &$1.07$&$6.0$& $2.793$ & $0.0116$&$0.27$\\ 
\hline
\end{tabular}
\end{table*}

The set of models can already be separated in three groups, with the somewhat younger subgiants being star A and B, C and D being in relatively intermediate stages and star E and F being already at the bottom of the RGB or starting to climb it. Due to the detailed modelling procedure carried out in \citet{Deheuvels2014}, we can consider that these models are suitable to compute rotation inversions. 

As all these stars are slow-rotators, the effects of rotation can be treated as a first order perturbation to the non-rotating solution. In this case, the so-called rotation splittings, $\delta \nu^{m}_{n,\ell}$,  are linked to the internal rotation profile, $\Omega$, by the following equation \citep{Hansen1977,Gough1981}
\begin{align}
\delta \nu^{m}_{n,\ell}=m\int_{0}^{R}K_{n,\ell}(r)\Omega(r)dr, \label{Eq:Splittings}
\end{align}
where we have assumed spherical symmetry, $R$ is the stellar radius and $K_{n,\ell}(r)$ are the rotation kernels. We neglect here the departure from spherical symmetry as it is unlikely to be measured from the limited set of data available here, although other studies \citep{Benomar2018,Hatta2019,Bazot2019} have mentioned it in main-sequence stars. The rotation splittings are taken from Tables 3-8 in \citet{Deheuvels2014}. 

\section{SOLA Rotation inversions}\label{Sec:SOLAInversions}

In their paper, \citet{Deheuvels2014} carried out MOLA inversions to determine the rotation of the stellar core. As a sanity check of the models and kernels, we verified that SOLA inversions provided the same results. The difference between MOLA and SOLA inversions resides in the cost function, which uses a substractive formulation of the term responsible for fitting the averaging kernel to the cost function instead of a multiplicative formulation. In practice SOLA should be slightly less efficient at localizing information but more stable and avoid oscillatory wings in the averaging kernels \citep{Pijpers,Sekii1997,Reese2018,Buldgen2022}. We briefly recall here the properties of the SOLA inversion. 

The SOLA cost function writes
\begin{align}
J(c_{k}(r_{t}))=&\int_{0}^{R}\left(T(r_{t},r) -K_{\rm{Avg}}(r_{t},r) \right)^{2} dr + \tan \theta \frac{\sum_{k=1}^{N} (c_{k}(r_{t})\sigma_{k})}{\langle\sigma^{2}\rangle}\nonumber \\ &+ \lambda \left( 1 - \int_{0}^{R} K_{\rm{Avg}}(r_{t},r) dr\right),
\end{align}

with $c_{k}(r_{t})$ the inversion coefficient associated with $r_{t}$, the target coordinate of the inversion, $T(r_{t},r)$ the target function of the inversion, $\theta$ a trade-off parameter to balance the contribution of the observed uncertainties with respect to the fit of the target function, $\lambda$ is a Lagrange multiplier used to ensure the normalisation of the averaging kernel, $K_{\rm{Avg}}(r_{t},r)$, defined as
\begin{align}
K_{\rm{Avg}}(r_{t},r)=\sum_{k=1}^{N}c_{k}(r_{t})K_{\Omega}(r),
\end{align}
and $\langle\sigma^{2}\rangle=\sum_{k=1}^{N}\frac{\sigma^{2}_{k}}{N}$. The target function is taken here as a Gaussian of the form
\begin{align}
T(r_{t},r)=Ar \exp\left(-\left(\frac{r-r_{t}}{\Delta c(r_{t})}-\frac{\Delta c(r_{t})}{2 r_{t}}\right)^{2} \right),
\end{align}
with $\Delta$ the width of the Gaussian that is a free parameter and $c(r_{t})$ the adiabatic sound speed at the target point of the inversion. This form of target function was  defined in \citet{RabelloParam} in the context of helioseismic inversions. 

\subsection{Application to \textit{Kepler} subgiants}\label{Sec:SOLAApplications}

As discussed in \citet{Sekii1997}, OLA and RLS inversions have a complementary way of analysing the data, and therefore their agreement can be seen as a sign of robustness of the inversion result. Both RLS and MOLA were used in \citet{Deheuvels2014} to determine the rotation of the subgiants studied here. We chose to use SOLA with $r_{t}$ placed at $0.01$R and $0.99$R, with R the photospheric radius. Given the limited set of splittings and the low degree of the modes, it would be overly optimistic to attempt an inversion of intermediate layers and we restrict ourselves to the core and outermost layers. As noted by \citet{Deheuvels2014}, there can be a pollution of the results of the envelope by the core, but this was measured to be negligible, with the exception of star F. For star E, there are two splittings that were not used for the inversion, namely those at 670 and 698 $\mu$Hz as they were deemed unreliable by \citet{Deheuvels2014}.

\begin{figure}
	\centering
		\includegraphics[width=9cm]{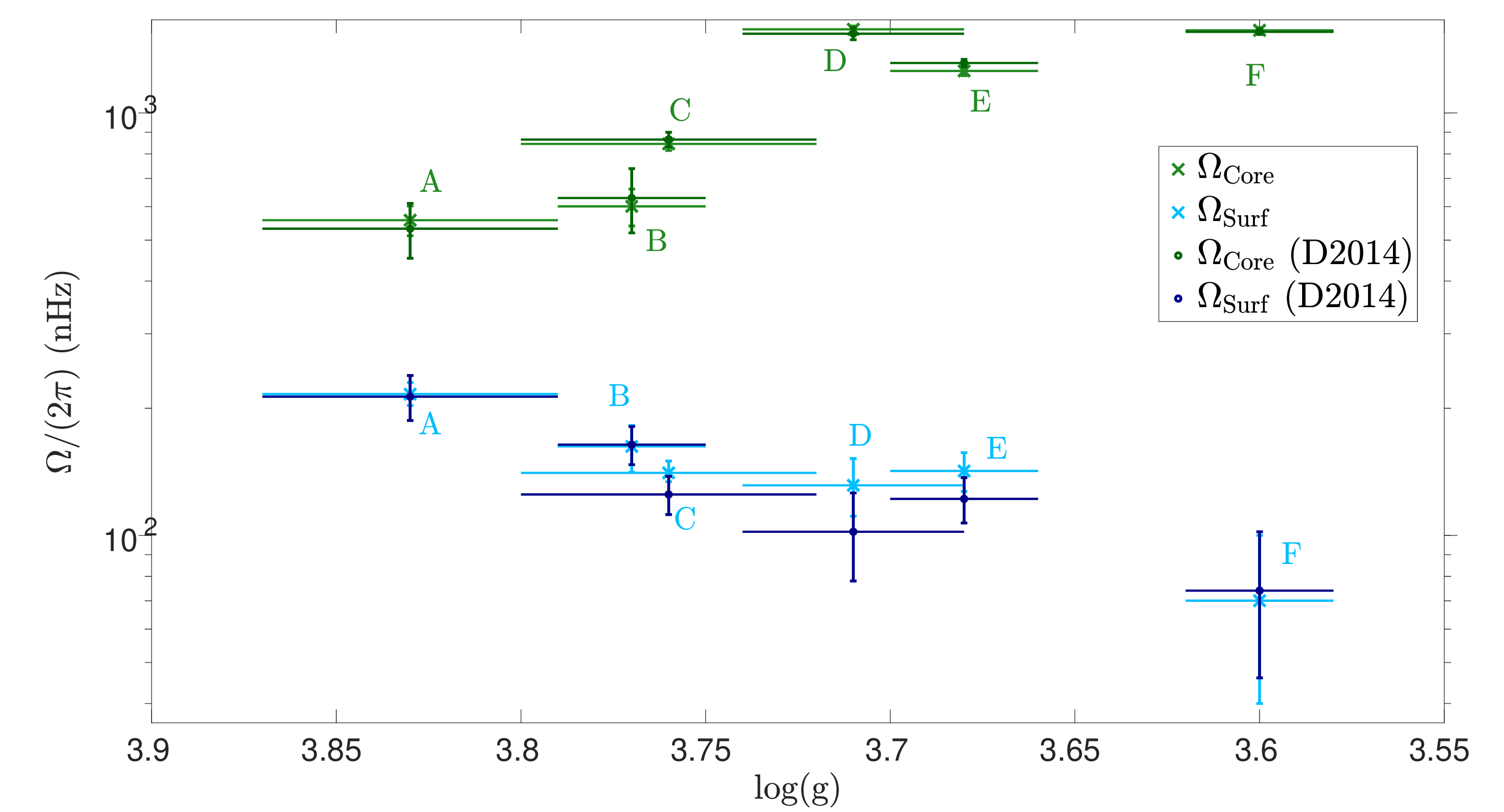}
	\caption{SOLA inversions results for the core and surface rotations of all stars in the \textit{Kepler} set.}
		\label{Fig:SOLA}
\end{figure} 

\begin{figure}
	\centering
		\includegraphics[width=9cm]{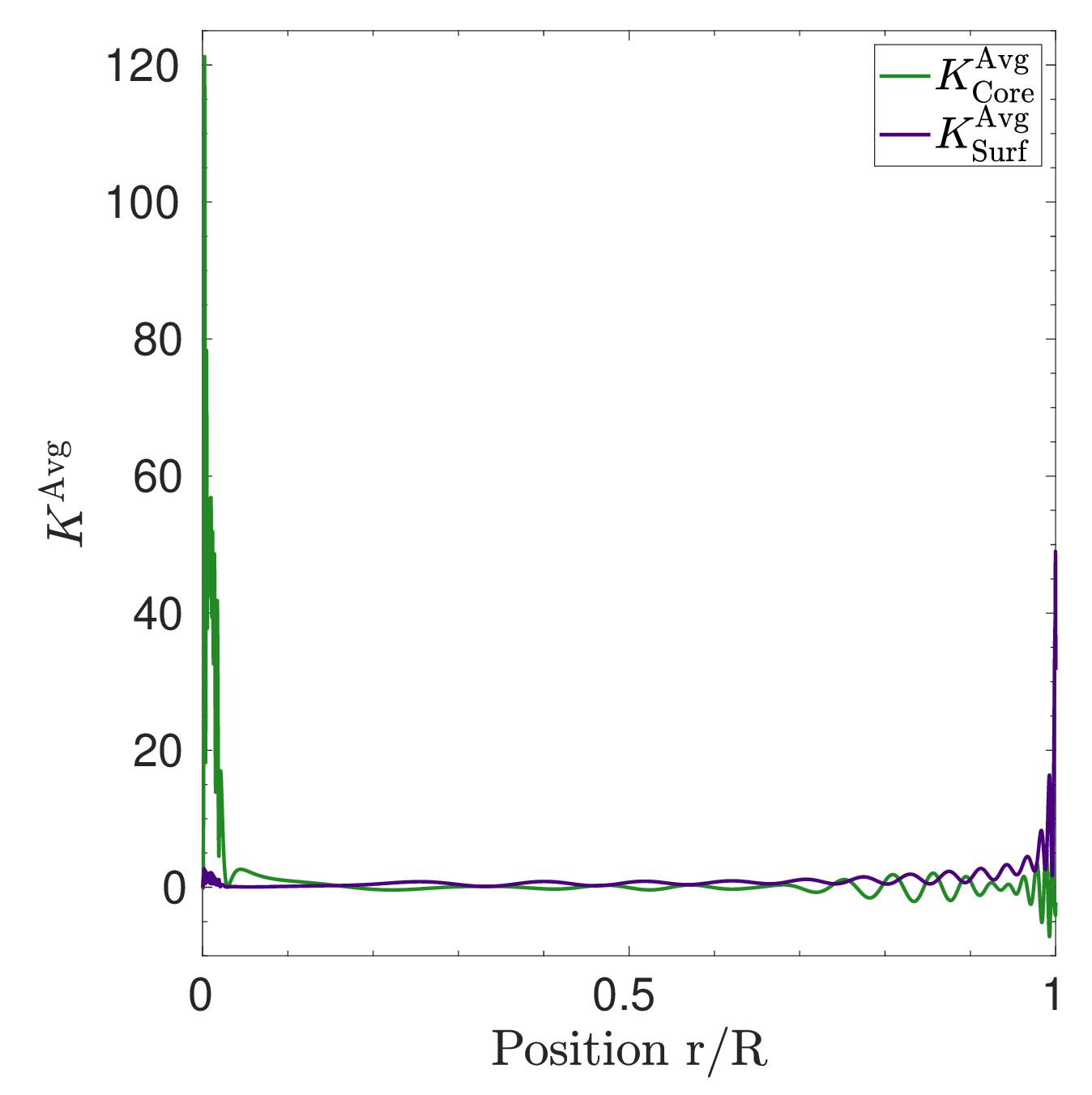}
	\caption{Averaging kernel for the case of Star C for the core rotation (green) and surface rotation (purple) as a function of the normalized radius.}
		\label{Fig:SOLAKer}
\end{figure}

Our results are illustrated in Fig.\ref{Fig:SOLA} and are compatible with the findings of \citet{Deheuvels2014} using MOLA. We note however that the SOLA inversion was in some cases more unstable and could lead to slightly different results depending on the chosen width of the target function and trade-off parameters for the errors. This was associated with a strong oscillatory behaviour in the core layers, underlining the sensitivity of the surface rotation inference to that of the core. We illustrate in Fig. \ref{Fig:SOLAKer} the averaging kernels for the inner and outermost layers of Star C, as can be seen, some residual oscillations are present at the source when inferring the core rotation, while some signal also remains in the core when inferring the surface rotation. This underlines the difficulties of using such a limited set of splittings, which may lead to some biases in the inversion results, as well as high sensitivity to the trade-off parameters since the illustrated oscillations might vary in amplitude rapidly when changing these parameters.

\section{MCMC Rotation inversions}\label{Sec:MCMC}
The background of MCMC inversions was first laid out by \citet{Fellay2021} with an application to Kepler 56 and independently applied by \citet{Hatta2022} to KIC11145123 . As mentioned above, \citet{Wilson2023} have applied this approach to Star A of the set, including surface rotation constraints from \citet{Garcia2014} for this particular star. They noted that they favour a transition around $0.2$ stellar radii when including this additional information. However, as can already be seen from Fig. 7 from \citet{Deheuvels2014}, Star A is a difficult case to attempt such inversions as the precision on the observed splittings is quite low, unfortunately. Additionaly, \citet{Wilson2023} did not test multiple forms of the internal rotation profile and only tested a two-zone model. 

\subsection{Theoretical framework and application to artifical data}\label{Sec:MCMCTheory}

The MCMC inversion is based on the approach developed in \citet{Fellay2021}, where the Monte Carlo Markov Chain algorithm is used to optimize the free parameters of parametric rotation profiles injected in Eq. \ref{Eq:Splittings}. For each walker, at each step, the splittings are evaluated using Eq. \ref{Eq:Splittings} and the likelihood of the model computed following 
\begin{align}
\mathcal{L}=-\frac{1}{2}\chi^{2},
\end{align}
with 
\begin{align}
\chi^{2}=\sum_{k=1}^{N}\left(\frac{\delta \nu_{Th,k} - \delta \nu_{Obs,k}}{\sigma_{k}}\right)^{2},
\end{align}
with $\delta \nu_{Obs,k}$ the observed rotation splitting, $\delta \nu_{Th,k}$ the theoretical splitting generated from the parametric profile for a given set of parameters, $\sigma_{k}$ the one sigma uncertainty on the observed splittings. 

We then compute the Bayesian Information Criterion (BIC)
\begin{align}
\mathrm{BIC}=k \rm{ln}(n)-2\mathcal{\hat{\mathcal{L}}}
\end{align}
with $k$ the number of free parameters, $n$ the number of constraints and $\hat{\mathcal{L}}$ the maximum of the likelihood.

We used 40 walkers with 2500 iterations, after 300 burn-in iterations. Given the high level of degeneracy in the parameter space, we used parallel tempering with 6 temperatures \citep[see e.g.][for an introduction to MCMC techniques]{Gelman2013}. The results correspond to the median of the distributions of the parameters and the errors are the first 15.9$\%$ and 84.1$\%$. We follow the guidelines of \citet{Foreman2013} that the number of iterations should be at least 50 times larger than the autocorrelation of the results for all runs presented here.

We start by presenting the five parametric rotation profiles we considered in our study.  We assume uniform priors on all free parameters of these profiles. For the core and surface rotations, we assume uniform priors centered around the values from the SOLA inversions allowing for $4\sigma$ variations (with the surface rotation always assumed positive, therefore rejecting counter-rotating solutions for the surface). For parameters denoting transitions in rotation, we enforce that the transition is located between $\left[0,1\right]$ in normalized radius. For the power law parameters, denoted $\alpha$, we enforce it to be between $\left[0, 40\right]$. We recall that theoretical predictions for this parameter suggest a value between $1$ and $1.5$. From their mathematical definitions, it is clear that some more complex parametrizations may degenerate into simpler ones for values of the parameters. 

\subsubsection{Step rotation profile}

The first rotation profile considered is a simple two zone model, defined by 
\begin{align}
\Omega(r)&=\Omega_{c}, \;\;\; (\rm{if}\; r<10^{\rm{t}}) \nonumber \\
\Omega(r)&=\Omega_{s}, \;\;\;
\end{align}
with $\Omega_{c}$, $\Omega_{s}$ and $\rm{t}$ the three free parameters of the profile characterizing the rotation in the inner layers, the rotation in the outer layers and the transition between the two zones, respectively. An exponent formulation for the parameter used to localize the transition is used here to avoid boundary effects in the distributions that lead to artefacts due to the physical constraint on $r_{t} > 0$ which may lead to walkers piling up at the border of the domain. 
\subsubsection{Gaussian profile}
The third profile is the Gaussian profile used in \citet{Fellay2021}, defined as
\begin{align}
\Omega(r)=(\Omega_{c}-\Omega_{s})\exp \left( -\left( r_{n}\times 10^{\rm{t}} \right)^{\beta} \right)+\Omega_{s},
\end{align}
with $r_{n}=0.02$ taken as a normalization constant linked with the peak of the Brunt-Väisälä, $\Omega_{c}$, the core rotation and $\Omega_{s}$ the surface rotation, $\sigma$ a multipicative factor assigning the location of the transition in rotation. $\beta$ is taken as a constant in the MCMC runs, fixed to 8 in \citet{Fellay2021} but in some of the following cases we changed its value to test smoother or sharper transitions. This profile is supposed to be able to mimic the transitions obtained from magnetic instabilities or at the very least angular momentum transport processes whose efficiency is inhibited by the presence of mean molecular weight gradients. This profile has three free parameters $\Omega_{c}$, $\Omega_{s}$ and $\rm{t}$.
\subsubsection{Power law profile}
The fourth profile is defined by
\begin{align}
\Omega(r)&=\Omega_{c}, \;\;\; (\rm{if}\; r<r_{BCE}) \nonumber \\
\Omega(r)&=\Omega_{c}\left( \frac{r_{BCE}}{r}\right)^{\alpha}, \nonumber \\
\label{Eq:PLaw}
\end{align}
with $r_{BCE}$ the position of the base of the convective envelope, $\omega_{c}$ the core rotation and $\alpha$ a free parameter constraining the decrease of rotation towards the surface. This profile has two free parameters. 
\subsubsection{Extended power law profile}
The fifth parametric profile is an extended version of the power law defined as
\begin{align}
\Omega(r)&=\Omega_{c}, \;\;\; (\rm{if}\; r<10^{\rm{t}}) \nonumber \\
\Omega(r)&=\Omega_{c}+\Omega_{t}\left(\left( \frac{10^{\rm{t}}}{r}\right)^{\alpha}-1 \right),
\end{align}
with $\Omega_{c}$ the core rotation, $\rm{t}$ linked with the location of the transition from the constant to the power law rotation profile, $\alpha$ describes the power law decline of the rotation profile and $\Omega_{t}$ is the difference between the core rotation $\Omega_{c}$ and the surface rotation that is reached in the upper layers. This rotation profile has 4  free parameters and is tailored to reach a non-zero surface rotation of $\Omega_{c}-\Omega_{t}$.
\subsubsection{Test cases on artificial data}
We started by testing the robustness of the inferences on artificial data. As already noted in \citet{Deheuvels2014},  not all the stars have equivalent data quality and it is also clear that a higher contrast in the profile leads to an easier fit. To check the potential and possible limitations of our method, we used the models of stars B, D and E, assumed a parametric rotation profile and checked the recovery capabilities of the MCMC, using the same set of rotation splittings and uncertainties as in the observed case. These tests thus serve as a good example of the capabilities of the technique and can be indicative of intrinsic limitations.

\begin{figure*}
	\centering
		\includegraphics[width=13cm]{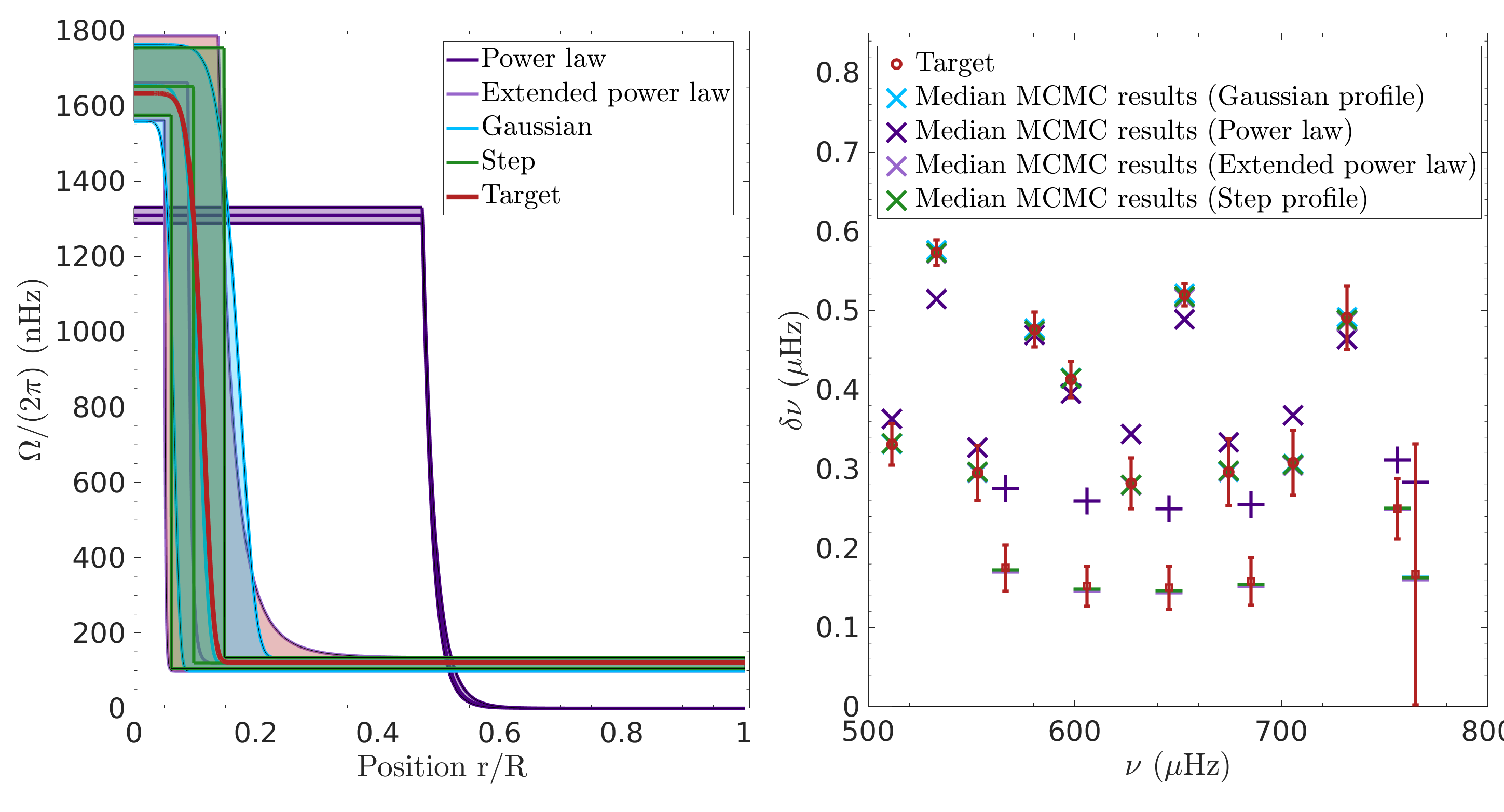}
	\caption{Recovery test for the MCMC method using the model and dataset of Star B. \textit{Left panel:} recovered rotation profiles as a function of normalized radius with the shaded area providing the $1\sigma$ uncertainties on the inferred profiles, the target being plotted in light purple. \textit{Right panel:} rotation splittings as a function of frequencies for the model of Star B (circles indicate dipolar modes, squares quadrupolar modes), using the simulated profile illustrated in light purple in the left panel.}
		\label{Fig:TestCaseB}
\end{figure*}

We illustrate the results for star B in Fig. \ref{Fig:TestCaseB}, where we have plotted in light purple the actual simulated rotation profile and in red, blue, green and dark purple the recovered median profiles from the MCMC. As can be seen, the splittings are well-fitted and the target profiles, indicated in purple, are quite well recovered. We note that attempts were made with for Star B with more complex rotation profiles (e.g. a three zone model) which led to rather poor results. It should also be noted that the rotation contrast is relatively high, which makes it easier to pick trends for the method. Nevertheless, this shows that Star B is already a promising target to attempt such MCMC inversions. We can also see that the power law profile does not at all recover the observed trends and leads to a poor fit of the rotation splittings. In other words, it is impossible for a power law profile to reproduce the splittings of a profile with a transition in the radiative layers as the shape of the assumed profile is incompatible with that of the target. All other profiles are able to reproduce the splittings and recover the main trends in the rotation profile and provide an approximate location of the transition in rotation. 

\begin{figure*}
	\centering
		\includegraphics[width=13cm]{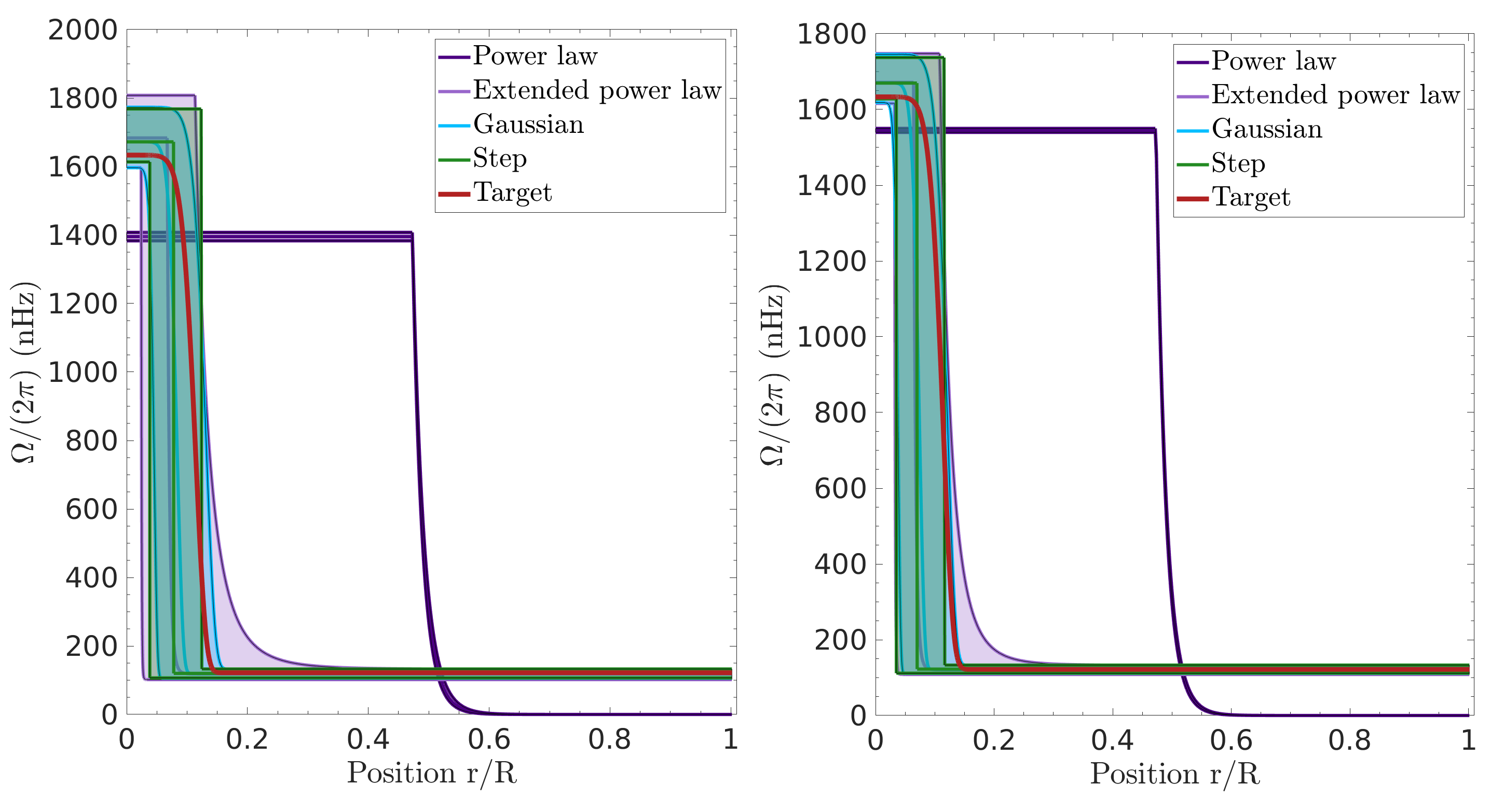}
	\caption{Recovery test for the MCMC method using the model and dataset of Star D and E. \textit{Left panel:} recovered rotation profiles as a function of normalized radius for Star D with the shaded area providing the $1\sigma$ uncertainties on the inferred profiles, the target being plotted in light purple. \textit{Right panel:} same for Star E.}
		\label{Fig:TestCaseDE}
\end{figure*}

\begin{figure}
	\centering
		\includegraphics[width=7cm]{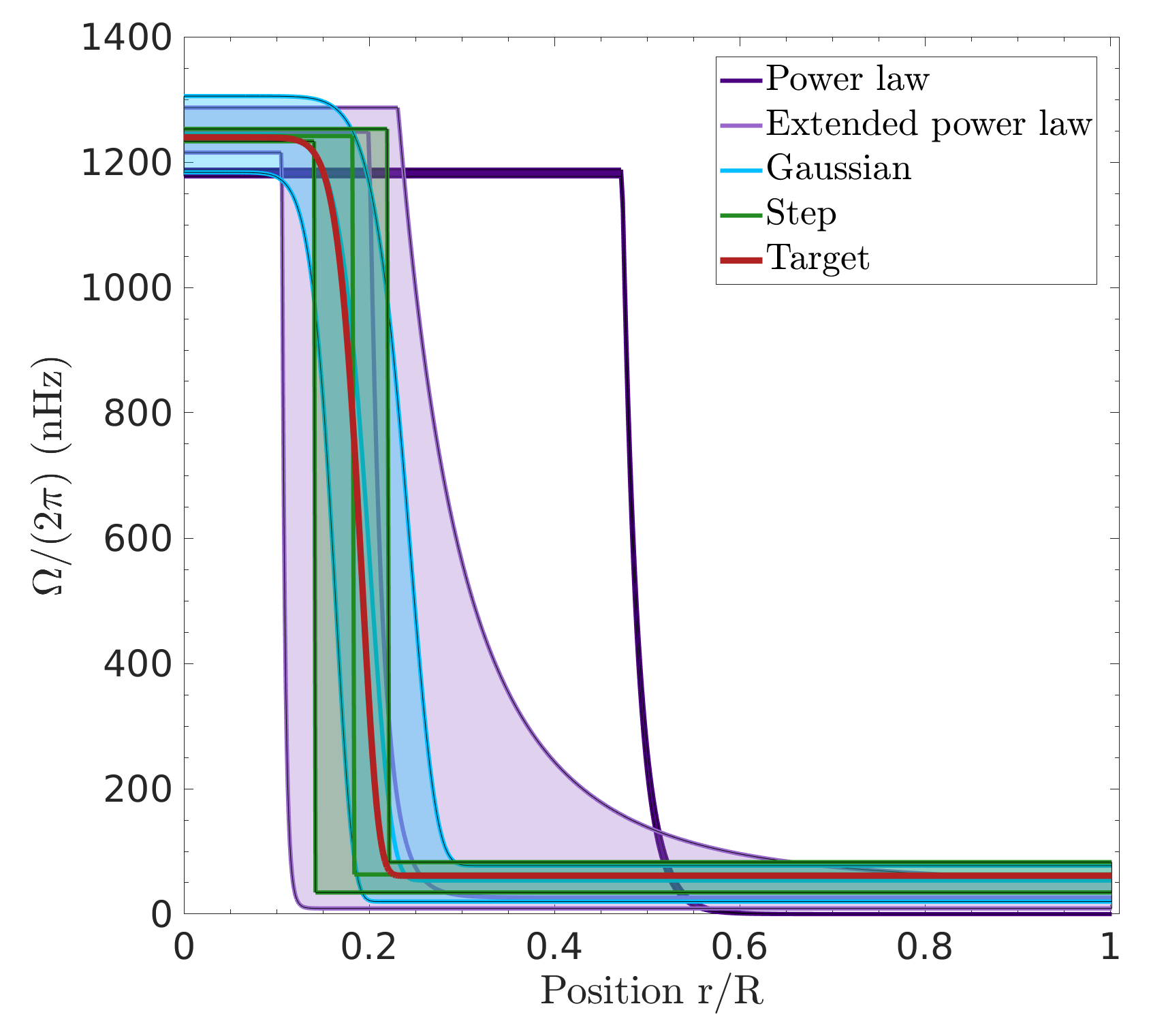}
	\caption{Recovery test for the MCMC method using the model and dataset of Star E using a target profile with a transition in rotation located higher in the star.}
		\label{Fig:TestCaseEHigh}
\end{figure}

The situation is similar for stars D and E, as illustrated in Fig. \ref{Fig:TestCaseDE}. In both cases, the method is able to pick up the trends of the profile. Even in cases where analytical expression of the actual target profile and the assumed one for the MCMC are significantly different, the localization of the transition is relatively accurate. This gives us confidence that some useful information about the properties of the rotation profile can actually be extracted, keeping in mind that there is an intrinsic limitation of how many free parameters can be used to describe the profile given the limited amount of data. For the sake of completeness, we also tested in the case of Star E a simulated profile with a transition in rotation located at a higher position in the star, this is illustrated in Fig. \ref{Fig:TestCaseEHigh}.
One important outcome is that a high transition uncorrelated with the peak of the Brunt-Väisälä frequency is clearly distinguished from a deep transition close to the region of strong chemical gradients. 

\subsection{Application to observed data and constraints on the internal rotation profile}\label{Sec:MCMCOBS}

\begin{figure*}
	\centering
		\includegraphics[width=13cm]{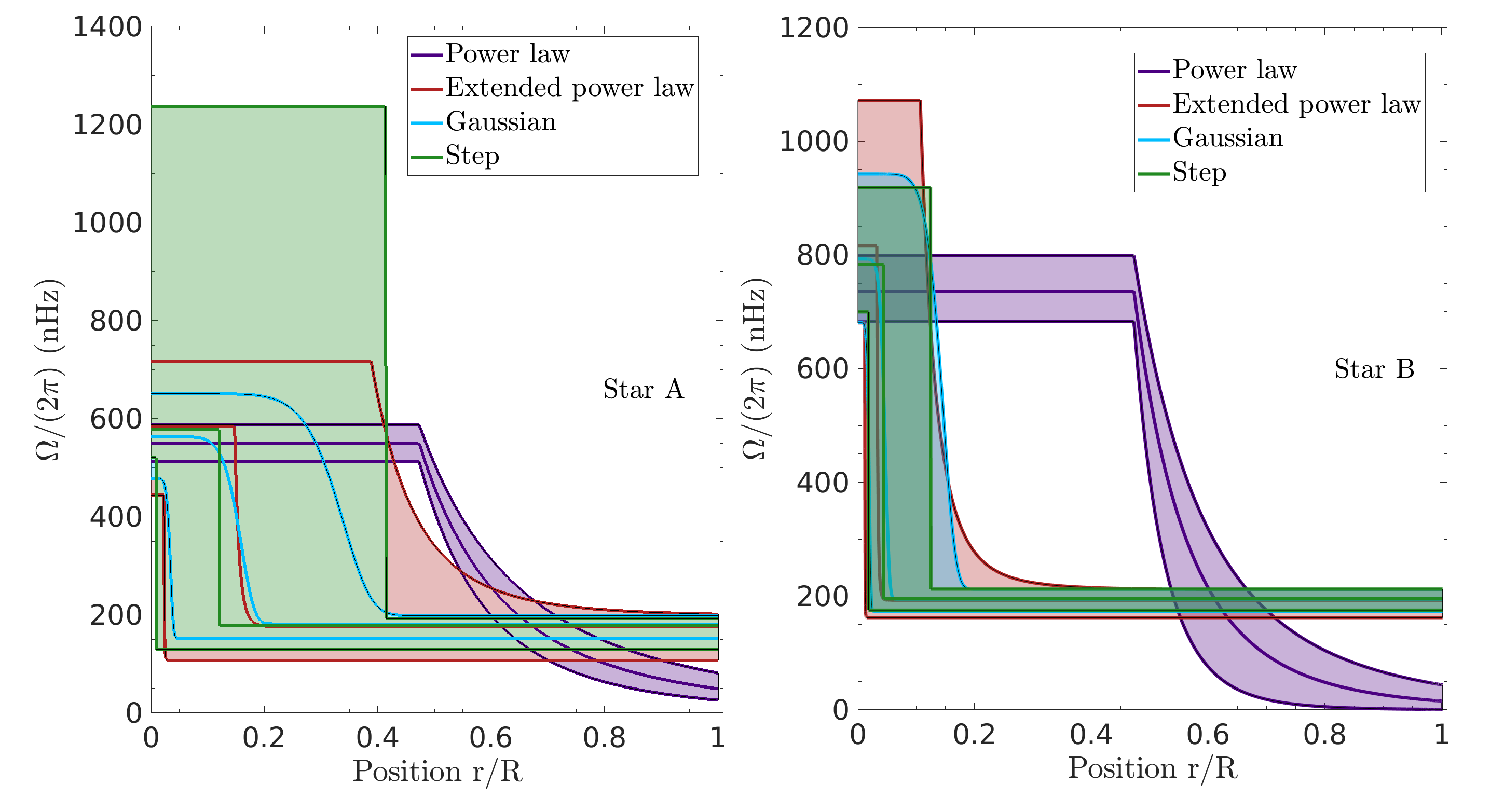}
	\caption{Inferred rotation profile as a function of normalized radius for Star A and B using the MCMC inversion technique and the various parametrized profiles of Sect. \ref{Sec:MCMCTheory}. The shaded areas show $1\sigma$ uncertainties on the inferred profiles. \textit{Left panel:} case of Star A. \textit{Right panel:} case of Star B.}
		\label{Fig:PanelMCMCRota1}
		\includegraphics[width=13cm]{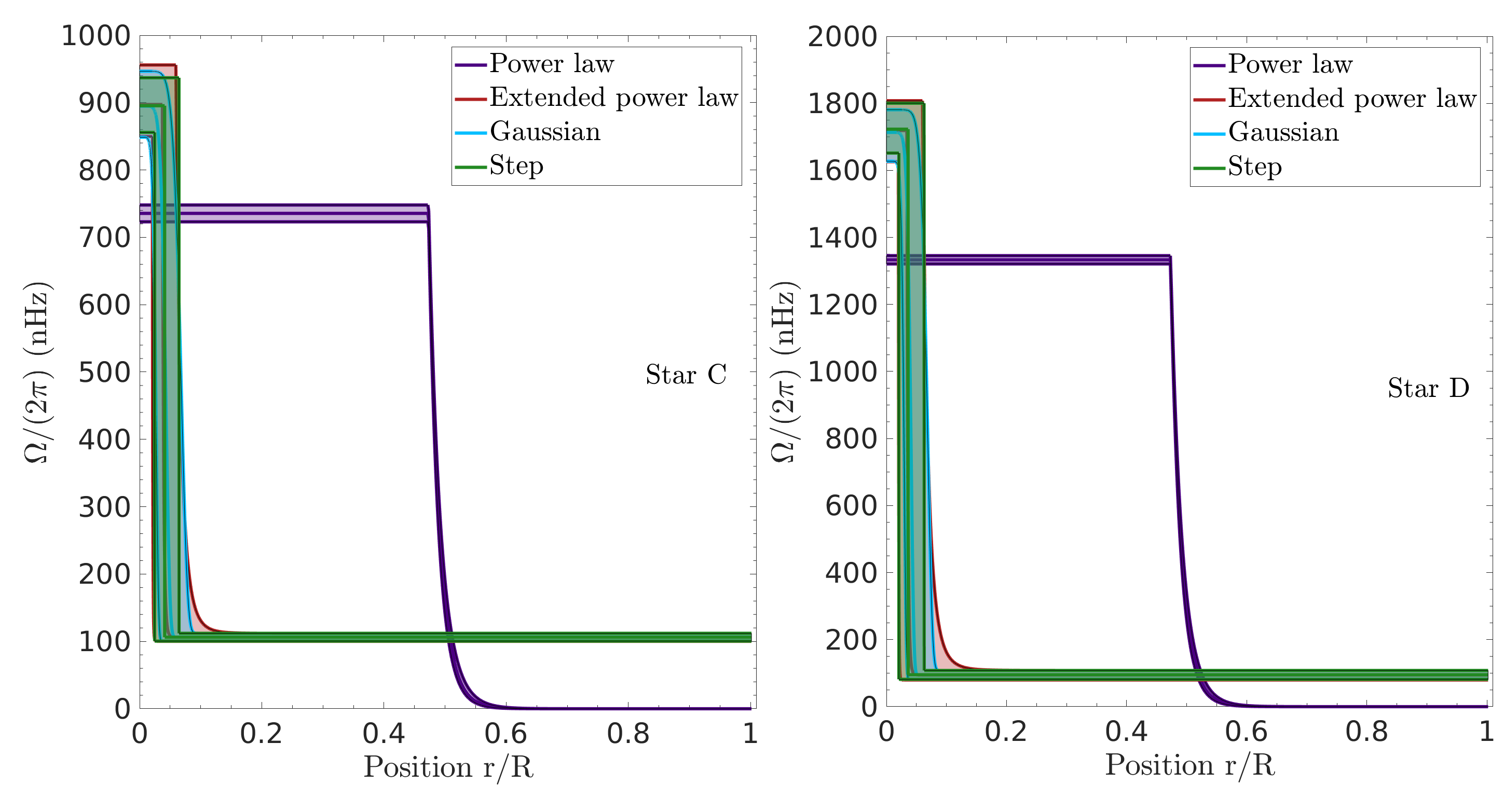}
	\caption{Same as Fig. \ref{Fig:PanelMCMCRota1} for Star C (left panel) and D (right panel).}
		\label{Fig:PanelMCMCRota2}
		\includegraphics[width=13cm]{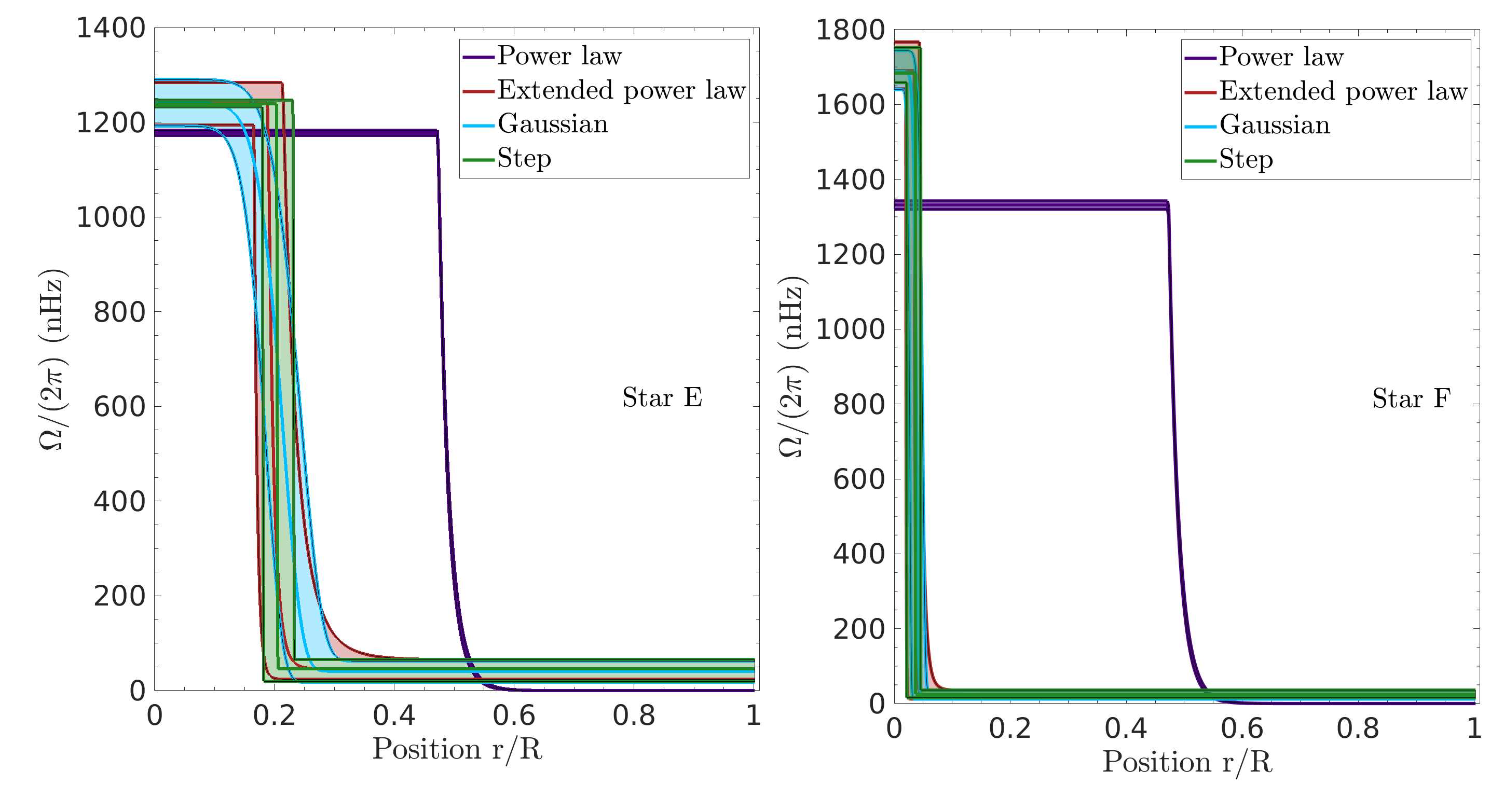}
	\caption{Same as Fig. \ref{Fig:PanelMCMCRota1} for Star E (left panel) and F (right panel).}
		\label{Fig:PanelMCMCRota3}
\end{figure*}

\begin{figure*}
	\centering
		\includegraphics[width=13cm]{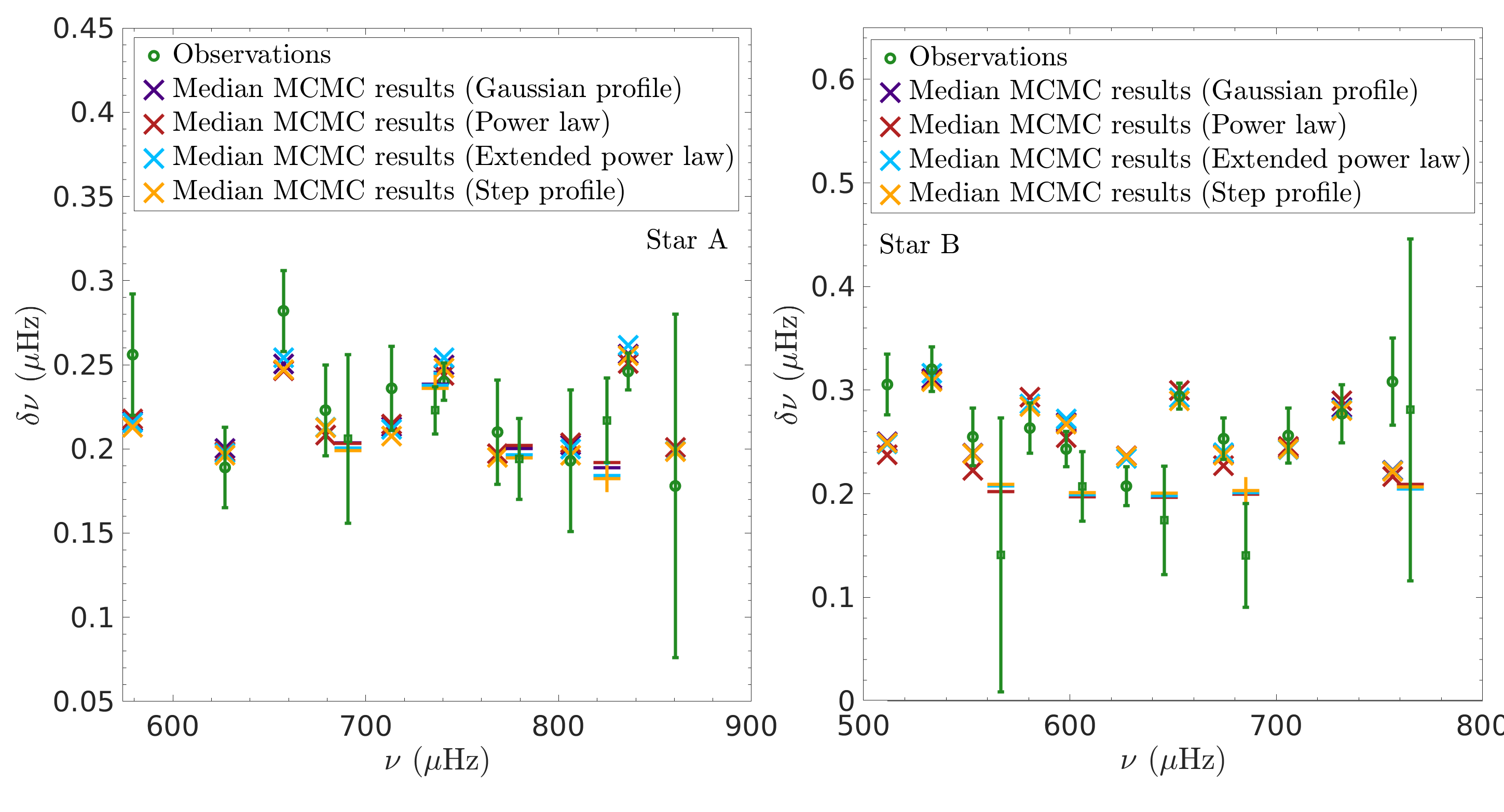}
	\caption{Agreement between the inferred rotation splittings for the various median parametric rotation profile and the observations (green) for Stars A and B (circles indicate dipolar modes, squares quadrupolar modes). \textit{Left panel:} results Star A. \textit{Right panel:} results for Star B.}
		\label{Fig:PanelMCMCSplittings1}
	\centering
		\includegraphics[width=13cm]{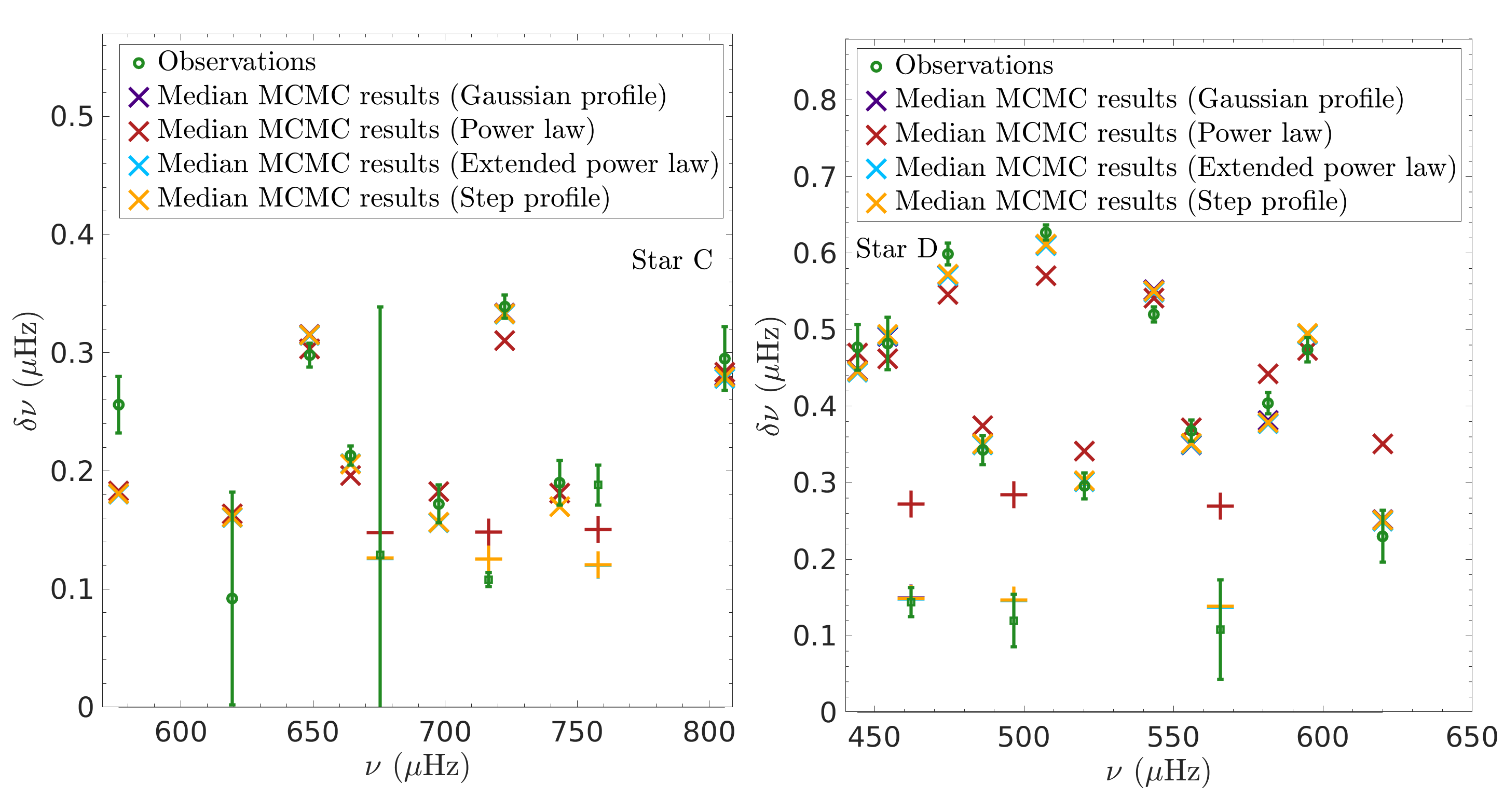}
	\caption{Same as Fig. \ref{Fig:PanelMCMCSplittings1} for Star C (left panel) and D (right panel).}
		\label{Fig:PanelMCMCSplittings2}
	\centering
		\includegraphics[width=13cm]{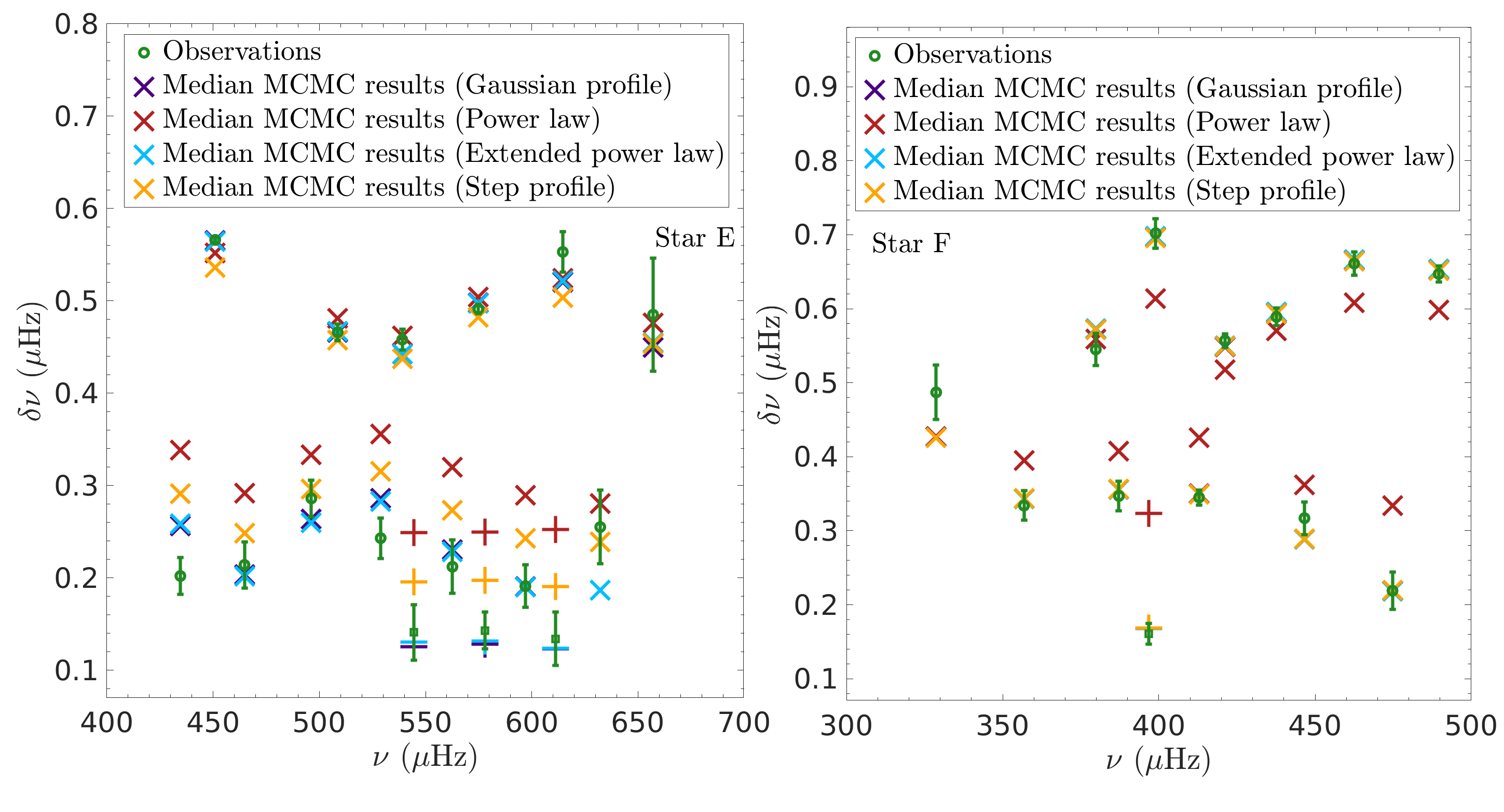}
	\caption{Same as Fig. \ref{Fig:PanelMCMCSplittings1} for Star E (left panel) and F (right panel).}
		\label{Fig:PanelMCMCSplittings3}
\end{figure*}

From the previous section, we have seen that the MCMC inversion could provide meaningful constraints on the shape of the internal rotation of the \textit{Kepler} subgiants. Although the case of younger stars seems more difficult, stars D, E and F still seem promising. We summarize in Figs. \ref{Fig:PanelMCMCRota1}, \ref{Fig:PanelMCMCRota2}, \ref{Fig:PanelMCMCRota3} the results of the MCMC inversion for each star, while Figs. \ref{Fig:PanelMCMCSplittings1}, \ref{Fig:PanelMCMCSplittings2}, \ref{Fig:PanelMCMCSplittings3} show the agreement with the observed splittings we reached. Some distributions for the parameters of the rotation profile are illustrated in the Appendix \ref{sec:Appendix} for the Gaussian, Step and Extended Power Law profile and Table \ref{tabSummary} summarizes our results. 

\def\arraystretch{1.5}

\begin{table}[]
\caption{Median parameter values determined for each star from our MCMC inversion.}
\label{tabSummary}
\resizebox{\hsize}{!}{\begin{tabular}{llllllll}

\hline\hline

Name     &            & Star A & Star B & Star C & Star D & Star E & Star F \\ \hline

         & $\rm{t}$   & $0.91^{+0.32}_{-0.66}$ &   $0.44^{+0.32}_{-0.28}$ & $0.32^{+0.28}_{-0.29}$ & $0.32^{+0.22}_{-0.15}$ & $1.01^{+0.10}_{-0.09}$ & $0.29^{+0.11}_{-0.16}$ \\
Gaussian & $\Omega_s$ & $181^{+17}_{-29}$ & $175^{+14}_{-20}$ & $118^{+6}_{-6}$ & $95^{+12}_{-14}$ & $55^{+23}_{-38}$ & $24^{+12}_{-12}$ \\

         & $\Omega_c$ & $563^{+70}_{-56}$ & $825^{+75}_{-68}$   & $877^{+78}_{-45}$ & $1618^{+54}_{-73}$ & $1185^{+27}_{-27}$ & $1664^{+45}_{-37}$\\ \cline{2-8} 

         & $\rm{BIC}$        &   $16.59$      &   $25.27$      &    $48.57$     &   $33.13$      &   $31.63$      &    $16.09$    \\ \hline

         & $\rm{t}$      & $-0.92^{+0.54}_{-1.12}$ & $-1.46^{+0.45}_{-0.53}$ & $-1.43^{+0.30}_{-0.20}$ & $-1.44^{+0.24}_{-0.24}$ & $-0.74^{+0.06}_{-0.06}$ & $-1.43^{+0.10}_{-0.23}$ \\

Step     & $\Omega_s$ & $177^{+15}_{-48}$ & $176^{+13}_{-17}$ & $123^{+6}_{-7}$ & $95^{+13}_{-13}$ & $65^{+17}_{-28}$ & $25^{+11}_{-9}$ \\

         & $\Omega_c$ & $577^{+646}_{-57}$ & $861^{+160}_{-76}$ & $892^{+50}_{-43}$ & $1723^{+77}_{-71}$ & $1237^{+8}_{-8}$ & $1684^{+67}_{-26}$ \\ \cline{2-8} 

         & $\rm{BIC}$        &      $14.69$   &    $22.45$    &  $45.91$     &     $30.21$   &   $28.49$   &  $13.10$  \\ \hline

Power  & $\alpha$ & $3.22^{+0.76}_{-0.58}$ &  $10.8^{+12}_{-3.3}$ & $26.2^{+2.7}_{-4.4}$ & $28.2^{+2.0}_{-3.0}$ & $28.7^{+1.0}_{-2.0}$ &    $29^{+0.7}_{-1.5}$ \\

Law      & $\Omega_c$ & $551^{+71}_{-38}$ & $795^{+33}_{-45}$ & $746^{+12}_{-13}$ & $1333^{+12}_{-13}$ & $1180^{+6}_{-6}$ & $1331^{+11}_{-11}$ \\ \cline{2-8} 

         & $\rm{BIC}$        &      $15.14$   &    $28.55$     &  $48.34$      &    $164.55$    &   $220.27$   &   $326.05$ \\ \hline

          & $r_t$ & $-0.83^{+0.42}_{-0.81}$ & $-1.43^{+0.45}_{-0.47}$ & $-1.53^{+0.28}_{-0.25}$ & $-1.47^{+0.24}_{-0.24}$ & $-0.75^{+0.10}_{-0.08}$ & $-1.46^{+0.10}_{-0.23}$ \\

Power     & $\alpha$ & $19^{+14}_{-14}$ & $18^{+15}_{-14}$ & $-20.0^{+13}_{-14}$ & $-20.0^{+13}_{-14}$ & $-20.0^{+14}_{-13}$ & $-21.0^{+14}_{-12}$ \\

Law     & $\Omega_s$ & $176^{+21}_{-70}$ & $113^{+32}_{-71}$ & $123^{+6}_{-8}$ & $94^{+15}_{-14}$ & $58^{+21}_{-37}$ & $23^{+12}_{-11}$ \\

Var           & $\Omega_c$ & $407^{+113}_{-70}$ & $680^{+214}_{-82}$ & $781^{+100}_{-49}$ & $1624^{+76}_{-78}$ & $1181^{+40}_{-16}$ & $1667^{+54}_{-38}$ \\ \cline{2-8} 

         & $\rm{BIC}$        &      $21.12$   &    $28.29$    &     $50.74$       &    $35.65$     &     $34.20$     &  $18.52$ \\ \hline

\end{tabular}}
\end{table}

\def\arraystretch{1.0}

A clear result in all cases is that the rotation profile following the expected properties of large-scale fossil fields as in \citet{Kissin2015} (a power law style profile with $\alpha \in \left[ 1.0,1.5 \right]$) is not providing a good agreement with the rotation splittings. This is directly seen from the BIC values in Table \ref{tabSummary} for stars D, E and F. For stars A, B and C, a lower BIC is reached but for values of the $\alpha$ parameter controlling the stiffness of the power law much higher than $1.5$, well above the allowed theoretical range from \citet{Kissin2015}. These results show the importance of taking into account a detailed description of the internal rotation profile and directly extracting the rotation splitting for a given star. Indeed, \citet{Takahashi2021} showed that they could provide a relatively good agreement with the results from \citet{Deheuvels2014} when trying to reproduce the core and surface rotation of the stars (see their Fig. 12). Here, we show that directly using the models reproducing the full-oscillation spectrum and simulating the rotational splittings, the agreement is actually bad. If we keep values of $\alpha$ between $1.0$ and $1.5$ as prescribed from a theoretical point of view the fitting of the splitting is very poor for all stars. If we do not restrain the values of $\alpha$ in Eq. \ref{Eq:PLaw}, the profile always converges on a strong transition and very slow, unphysical, surface rotation, while still not providing an adequate fit for the splittings of the most p-dominated modes. This implies that the transition is likely below the base of the convective zone, in the radiative interior, to allow to reproduce the splittings of the p-dominated modes without requiring extremely low surface rotation values.

The second conclusion that we can draw is that a deep transition is not clearly favoured in the young subgiants either, but this is to be taken cautiously since its localization is quite challenging. It should also be noted that star B shows some unusual behaviour in the splittings, with some dipolar modes showing higher splitting values than the quadrupolar ones even at a similar coupling level, this leads ot some difficulties in reproducing the splittings.Such behaviour is not observed in other stars, although some splittings in Star C and D show a similar behaviour. It is unclear yet whether this is a result of the forward modelling procedure used to determine the properties of the evolutionary model, which could affect the coupling, or of some physical behaviour in the rotation profile (latitudinal differential rotation, more slowly rotating core, ...) which the inversion is unable to pick. 

\begin{figure*}
	\centering
		\includegraphics[width=13cm]{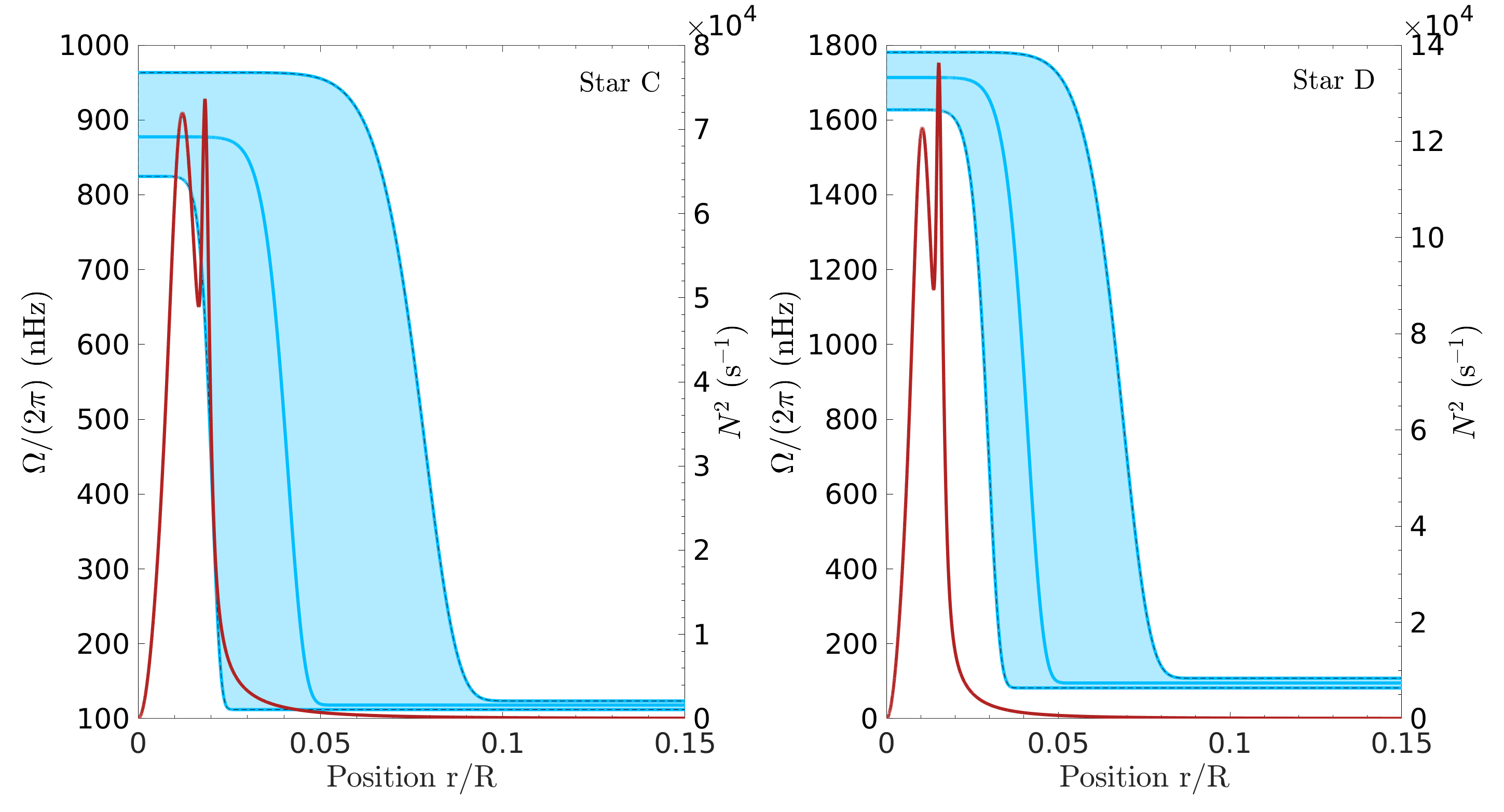}
	\caption{Comparison of the location of the transition in rotation (left y-axis) with the Brunt-Väisälä frequency profile (right y-axis) for Stars C (left panel) and D (right panel).}
		\label{Fig:PanelN2Rota1}
	\centering
		\includegraphics[width=13cm]{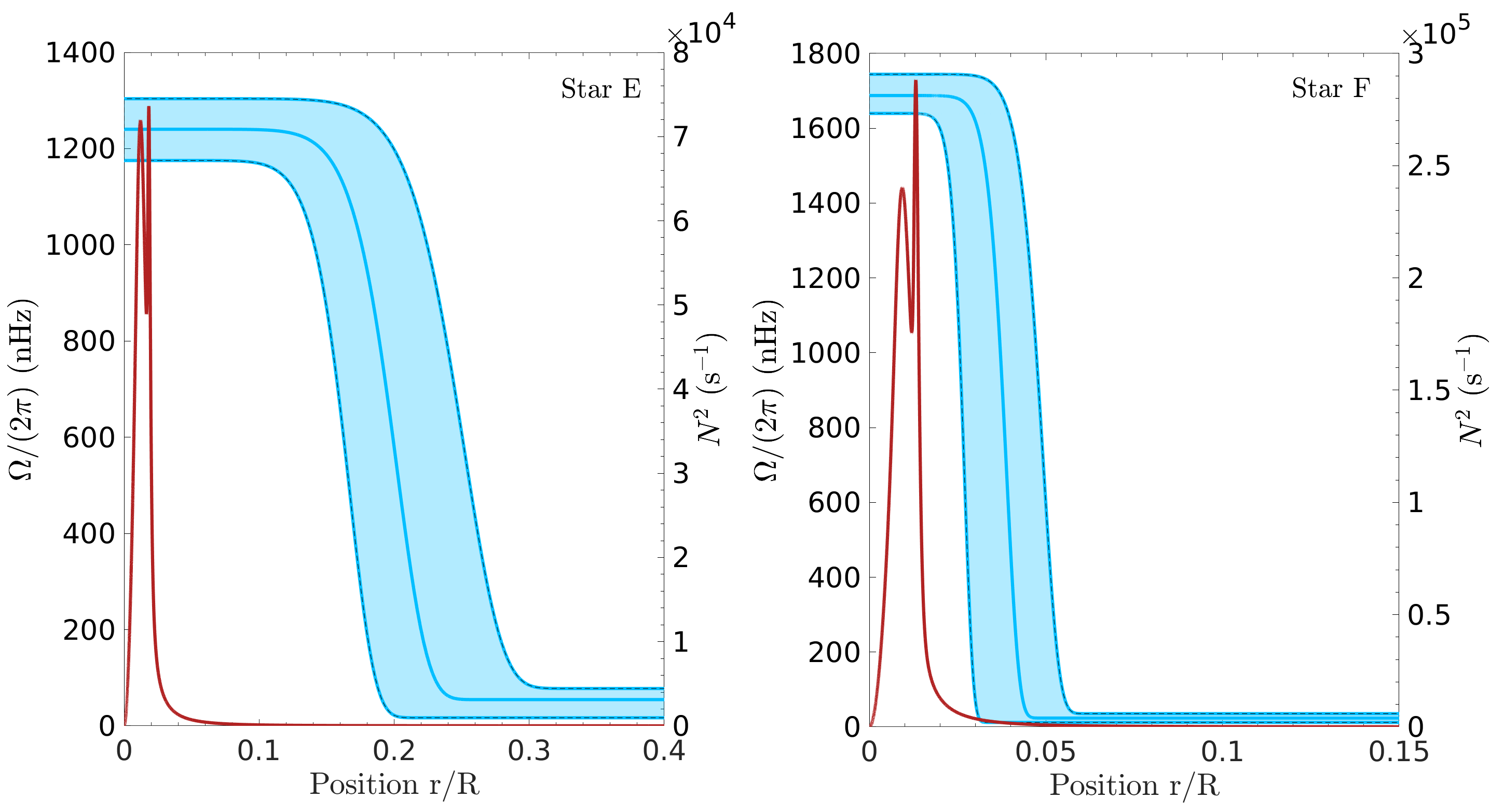}
	\caption{Same as Fig. \ref{Fig:PanelN2Rota1} but for stars E and F}
		\label{Fig:PanelN2Rota2}
\end{figure*}

The more evolved subgiants show some more interesting features, with the lower mass range (stars C, D and F) showing a location of the rotation transition close to (but not exactly at) the position of the peak in the Brunt-Väisälä frequency, in line with previous results \citep{DiMauro2016, DiMauro2018, Fellay2021}. This is illustrated in Figs. \ref{Fig:PanelN2Rota1}, \ref{Fig:PanelN2Rota2} which show the position of the median profile and the best fit profile against the Brunt-Väisälä frequency profile for Stars C, D, E and F. Namely, the transition in rotation is linked with the position in the star where the effects of mean-molecular weight gradients seem to dominate. This is in line with both the effects of magnetic instabilities and internal gravity waves. Star E however, the most massive of the sample, shows a transition in rotation clearly unrelated to the location of the chemical composition gradient, in fact much higher in relative radius than any other star. We emphasize that these conclusions can be drawn from multiple parameterizations of the profile, some of them allowing for smooth transitions within the radiative zone. Such a smooth transition is not favoured in all cases, however it is unclear whether this is due to a lack of constraints in the splittings or an actual feature of the rotation profile. It is also clear that these are azimuthally averaged profiles and do not imply that some breaking from spherical symmetry is not occurring in these stars. 

\section{Discussion}\label{Sec:Disc}

From a closer look at the properties of the inferred profiles, we can try to draw a more general picture and discuss some additional perspectives. It is clear that our inversion results invalidate fossil magnetic fields in the whole radiative region as theorized by \citet{Kissin2015} and tested by \citet{Takahashi2021}. We thus confirm the results obtained by \citet{Fellay2021} using the same approach for Kepler 56 which invalidated such a shape for the internal rotation profile, even with a high value for the $\alpha$ parameter in the power law rotation profile. 

However, we can also say that it is still difficult to draw a full picture of the evolutionary trends in the rotation profile of the sample of subgiants from our results alone. For example, the case of stars A and B is somewhat problematic, with difficulties in localizing the transition in rotation, although hinting at a transition disconnected from the peak in the Brunt-Väisälä frequency or a smoother transition over an extended region. We also note some difficulties in reproducing all splittings for the case of star B. For star C, a better localization is achieved, although two splittings seem to be difficult to reproduce, perhaps also calling for a detailed remodelling of this star.

For the more evolved subgiants, a different picture can be drawn. It seems that the lower mass stars show a rotation transition linked with the chemical composition gradients, in line with the results of \citet{DiMauro2016}, \citet{DiMauro2018} and \citet{Fellay2021}. Star E, however, the more massive one, has a transition much higher, at about 28$\%$ of the total mass, while the maximum extension of the convective core on the main-sequence is around twice lower. Whether this implies the presence of magnetic fields generated in the convective core and remaining as a fossil fields in these layers is unclear and requires input from simulations regarding the survival of such fields and their confinement. This positioning of the transition is however robust and confirmed with various parametrizations as well as using artificial data.  

However, we have to remain cautious not to generalize these observations. Increasing the number of targets and carrying out a detailed modelling of both the stars in \citet{Deheuvels2014} and \citet{Deheuvels2020} is required to draw a clearer picture of the evolutionary trends. Star C seems to be favouring a transition close to the peak of Brunt-Väisälä, while star A does not. In that respect young subgiants appear to be the key to unravel the physical nature of the angular momentum transport during the late main sequence, as the key difference between magnetic instabilities and internal gravity waves being the rotation of the deep core. Indeed, internal gravity waves are expected to induce a more slowly rotating core, in line with \citet{Deheuvels2020}, while magnetic instabilities, even in their most efficient calibrated form \citep{Fuller2019} can enforce at best solid-body rotation. Combining this with detailed investigations of some red giant branch targets might also prove useful to provide a picture of the changes of the shape of the rotation profile as the star evolves and thus constrain the properties of the missing angular momentum transport.

A strong limitation of our inference results is the 1D structure of the profile. It is thus entirely possible than azimuthal patterns could be present and are not seen by the inversion procedure. Such effects need to be treated in a forward approach, with direct inputs from simulations averaged to determine the rotation splittings they would generate, in a similar fashion to what is done with entrainment laws derived from convective simulations. 

\section{Conclusion}\label{Sec:Conc}

In this study, we have analyzed in details the rotation properties of the 6 subgiants studied in \citet{Deheuvels2014}. We have confirmed their results based on MOLA and RLS inversion techniques by carrying out SOLA inversions and supplemented these approaches with MCMC inversions, following \citet{Fellay2021} and \citet{Hatta2022}. Using artificial data with the same uncertainties as the actual observations, we have demonstrated that our MCMC rotation inversion is able to recover some useful information about the internal rotation profile of these stars. We have thus shown that some major characteristics were accessible such as the core and surface rotation, as well as the location of the transition in rotation, using the same sets of modes and given uncertainties as in the observed cases, even if the profiles not exactly matched the target ones. We have also shown that there is a clear intrinsic limitation to the number of parameters that could be used to describe the profiles, namely around $\approx 3$ and that the younger stars in the sample were more difficult to work with. 

When applied to the observed targets, the MCMC inversion provides values compatible with the ones inferred with the SOLA inversion. Star C, D and F show a transition in rotation compatible with the position of the core transition, where the mean molecular weight gradients are expected to dominate the Brunt-Väisälä frequency profile. Star A and B do not show such clear signs but this could be due to the lower precision in the observed splittings for star A and to a difficulty to reproduce the observed splittings for star B. Whether this is due to an issue in the forward modelling procedure used to determine the evolutionary model or the parametrization being unsuitable requires further analysis. For star E, the transition in rotation is clearly localized and uncorrelated with the position of the mean molecular weight peak in the Brunt-Väisälä profile. Surprisingly, this star is also the most massive of the sample and the location of the transition is also of the same order of magnitude of the expected extension of convective core during the main-sequence. 

Another clear conclusion from our study is that the solution suggested by \citet{Kissin2015} and tested by \citet{Takahashi2021} is incompatible if a detailed comparison of the rotation splittings is made. This further confirms the findings of \citet{DiMauro2016}, \citet{DiMauro2018} and \citet{Fellay2021}. 

Our findings seem to indicate that a diverse behaviour may be expected to explain the rotation properties of subgiants and red giants. A consistent remodelling of the younger stars in the sample and a joint analysis of MCMC inversions including the young subgiants of \citet{Deheuvels2020} will also be very informative, as these stars seem to favour internal gravity waves as the efficient transport mechanism acting after the main-sequence \citep{Pincon2017}. Such a full re-analysis will be done in a future study. An increased set of targets for which such inferences are achievable would also be extremely constraining on the physical nature of the angular momentum transport process. This will be achieved with the future PLATO mission \citep{Rauer2014} which is expected to deliver high-quality seismic data for thousands of subgiant stars \citep{Goupil2024}. An important aspect of constraining angular momentum on the main-sequence would be the detection of gravity modes in F type stars \citep{Breton2023} as these would allow to fill a gap in our probing of internal rotation as \cite{Betrisey2023} demonstrated that p-modes are unable to infer the presence of an efficient AM transport process even for the best F-type \textit{Kepler} targets .

\section*{Acknowledgements}

We thank the anonymous referee for their careful reading of the manuscript. G.B. and J.B. are funded by the SNF AMBIZIONE grant No 185805 (Seismic inversions and modelling of transport processes in stars). G.B. acknowledges funding from  the Fonds de la Recherche Scientifique – FNRS. L.F. is supported by the Fonds de la Recherche Scientifique (FNRS) as a Research fellow.  S.D. acknowledges support from from the project BEAMING ANR-18-CE31-0001 of the French National Research Agency (ANR) and from the Centre National d’Etudes Spatiales (CNES). M.F. is a Postdoctoral Researcher of the Fonds de la Recherche Scientifique – FNRS. E.F. is support by SNF grant number $200020\_212124$.

\bibliography{biblioarticleSG}

\begin{thebibliography}{43}
\expandafter\ifx\csname natexlab\endcsname\relax\def\natexlab#1{#1}\fi

\bibitem[{{Aerts} {et~al.}(2019){Aerts}, {Mathis}, \& {Rogers}}]{Aerts2019}
{Aerts}, C., {Mathis}, S., \& {Rogers}, T.~M. 2019, \araa, 57, 35

\bibitem[{{Auvergne} {et~al.}(2009){Auvergne}, {Bodin}, {Boisnard}, {Buey},
  {Chaintreuil}, {Epstein}, {Jouret}, {Lam-Trong}, {Levacher}, {Magnan},
  {Perez}, {Plasson}, {Plesseria}, {Peter}, {Steller}, {Tiph{\`e}ne}, {Baglin},
  {Agogu{\'e}}, {Appourchaux}, {Barbet}, {Beaufort}, {Bellenger}, {Berlin},
  {Bernardi}, {Blouin}, {Boumier}, {Bonneau}, {Briet}, {Butler}, {Cautain},
  {Chiavassa}, {Costes}, {Cuvilho}, {Cunha-Parro}, {de Oliveira Fialho},
  {Decaudin}, {Defise}, {Djalal}, {Docclo}, {Drummond}, {Dupuis}, {Exil},
  {Faur{\'e}}, {Gaboriaud}, {Gamet}, {Gavalda}, {Grolleau}, {Gueguen},
  {Guivarc'h}, {Guterman}, {Hasiba}, {Huntzinger}, {Hustaix}, {Imbert},
  {Jeanville}, {Johlander}, {Jorda}, {Journoud}, {Karioty}, {Kerjean},
  {Lafond}, {Lapeyrere}, {Landiech}, {Larqu{\'e}}, {Laudet}, {Le Merrer},
  {Leporati}, {Leruyet}, {Levieuge}, {Llebaria}, {Martin}, {Mazy}, {Mesnager},
  {Michel}, {Moalic}, {Monjoin}, {Naudet}, {Neukirchner}, {Nguyen-Kim},
  {Ollivier}, {Orcesi}, {Ottacher}, {Oulali}, {Parisot}, {Perruchot},
  {Piacentino}, {Pinheiro da Silva}, {Platzer}, {Pontet}, {Pradines},
  {Quentin}, {Rohbeck}, {Rolland}, {Rollenhagen}, {Romagnan}, {Russ}, {Samadi},
  {Schmidt}, {Schwartz}, {Sebbag}, {Smit}, {Sunter}, {Tello}, {Toulouse},
  {Ulmer}, {Vandermarcq}, {Vergnault}, {Wallner}, {Waultier}, \&
  {Zanatta}}]{Auvergne2009}
{Auvergne}, M., {Bodin}, P., {Boisnard}, L., {et~al.} 2009, \aap, 506, 411

\bibitem[{{Bazot} {et~al.}(2019){Bazot}, {Benomar}, {Christensen-Dalsgaard},
  {Gizon}, {Hanasoge}, {Nielsen}, {Petit}, \& {Sreenivasan}}]{Bazot2019}
{Bazot}, M., {Benomar}, O., {Christensen-Dalsgaard}, J., {et~al.} 2019, \aap,
  623, A125

\bibitem[{{Beck} {et~al.}(2012){Beck}, {Montalban}, {Kallinger}, {De Ridder},
  {Aerts}, {Garc{\'\i}a}, {Hekker}, {Dupret}, {Mosser}, {Eggenberger},
  {Stello}, {Elsworth}, {Frandsen}, {Carrier}, {Hillen}, {Gruberbauer},
  {Christensen-Dalsgaard}, {Miglio}, {Valentini}, {Bedding}, {Kjeldsen},
  {Girouard}, {Hall}, \& {Ibrahim}}]{Beck2012}
{Beck}, P.~G., {Montalban}, J., {Kallinger}, T., {et~al.} 2012, \nat, 481, 55

\bibitem[{{Benomar} {et~al.}(2018){Benomar}, {Bazot}, {Nielsen}, {Gizon},
  {Sekii}, {Takata}, {Hotta}, {Hanasoge}, {Sreenivasan}, \&
  {Christensen-Dalsgaard}}]{Benomar2018}
{Benomar}, O., {Bazot}, M., {Nielsen}, M.~B., {et~al.} 2018, Science, 361, 1231

\bibitem[{{B{\'e}trisey} {et~al.}(2023){B{\'e}trisey}, {Eggenberger},
  {Buldgen}, {Benomar}, \& {Bazot}}]{Betrisey2023}
{B{\'e}trisey}, J., {Eggenberger}, P., {Buldgen}, G., {Benomar}, O., \&
  {Bazot}, M. 2023, \aap, 673, L11

\bibitem[{{Borucki} {et~al.}(2010){Borucki}, {Koch}, {Basri}, {Batalha},
  {Brown}, {Caldwell}, {Caldwell}, {Christensen-Dalsgaard}, {Cochran},
  {DeVore}, {Dunham}, {Dupree}, {Gautier}, {Geary}, {Gilliland}, {Gould},
  {Howell}, {Jenkins}, {Kondo}, {Latham}, {Marcy}, {Meibom}, {Kjeldsen},
  {Lissauer}, {Monet}, {Morrison}, {Sasselov}, {Tarter}, {Boss}, {Brownlee},
  {Owen}, {Buzasi}, {Charbonneau}, {Doyle}, {Fortney}, {Ford}, {Holman},
  {Seager}, {Steffen}, {Welsh}, {Rowe}, {Anderson}, {Buchhave}, {Ciardi},
  {Walkowicz}, {Sherry}, {Horch}, {Isaacson}, {Everett}, {Fischer}, {Torres},
  {Johnson}, {Endl}, {MacQueen}, {Bryson}, {Dotson}, {Haas}, {Kolodziejczak},
  {Van Cleve}, {Chandrasekaran}, {Twicken}, {Quintana}, {Clarke}, {Allen},
  {Li}, {Wu}, {Tenenbaum}, {Verner}, {Bruhweiler}, {Barnes}, \&
  {Prsa}}]{Borucki2010}
{Borucki}, W.~J., {Koch}, D., {Basri}, G., {et~al.} 2010, Science, 327, 977

\bibitem[{{Breton} {et~al.}(2023){Breton}, {Dhouib}, {Garc{\'\i}a}, {Brun},
  {Mathis}, {P{\'e}rez Hern{\'a}ndez}, {Mathur}, {Dyrek}, {Santos}, \&
  {Pall{\'e}}}]{Breton2023}
{Breton}, S.~N., {Dhouib}, H., {Garc{\'\i}a}, R.~A., {et~al.} 2023, \aap, 679,
  A104

\bibitem[{{Buldgen} {et~al.}(2022){Buldgen}, {B{\'e}trisey}, {Roxburgh},
  {Vorontsov}, \& {Reese}}]{Buldgen2022}
{Buldgen}, G., {B{\'e}trisey}, J., {Roxburgh}, I.~W., {Vorontsov}, S.~V., \&
  {Reese}, D.~R. 2022, Frontiers in Astronomy and Space Sciences, 9, 942373

\bibitem[{{Charbonnel} \& {Talon}(2005)}]{Charbonnel2005}
{Charbonnel}, C. \& {Talon}, S. 2005, Science, 309, 2189

\bibitem[{{Christensen-Dalsgaard}(2011)}]{JCDAdipls2011}
{Christensen-Dalsgaard}, J. 2011, {ADIPLS: Aarhus Adiabatic Oscillation Package
  (ADIPACK)}, Astrophysics Source Code Library, record ascl:1109.002

\bibitem[{{Deheuvels} {et~al.}(2015){Deheuvels}, {Ballot}, {Beck}, {Mosser},
  {{\O}stensen}, {Garc{\'\i}a}, \& {Goupil}}]{Deheuvels2015}
{Deheuvels}, S., {Ballot}, J., {Beck}, P.~G., {et~al.} 2015, \aap, 580, A96

\bibitem[{{Deheuvels} {et~al.}(2020){Deheuvels}, {Ballot}, {Eggenberger},
  {Spada}, {Noll}, \& {den Hartogh}}]{Deheuvels2020}
{Deheuvels}, S., {Ballot}, J., {Eggenberger}, P., {et~al.} 2020, \aap, 641,
  A117

\bibitem[{{Deheuvels} {et~al.}(2014){Deheuvels}, {Do{\u{g}}an}, {Goupil},
  {Appourchaux}, {Benomar}, {Bruntt}, {Campante}, {Casagrande}, {Ceillier},
  {Davies}, {De Cat}, {Fu}, {Garc{\'\i}a}, {Lobel}, {Mosser}, {Reese},
  {Regulo}, {Schou}, {Stahn}, {Thygesen}, {Yang}, {Chaplin},
  {Christensen-Dalsgaard}, {Eggenberger}, {Gizon}, {Mathis},
  {Molenda-{\.Z}akowicz}, \& {Pinsonneault}}]{Deheuvels2014}
{Deheuvels}, S., {Do{\u{g}}an}, G., {Goupil}, M.~J., {et~al.} 2014, \aap, 564,
  A27

\bibitem[{{Deheuvels} {et~al.}(2012){Deheuvels}, {Garc{\'\i}a}, {Chaplin},
  {Basu}, {Antia}, {Appourchaux}, {Benomar}, {Davies}, {Elsworth}, {Gizon},
  {Goupil}, {Reese}, {Regulo}, {Schou}, {Stahn}, {Casagrande},
  {Christensen-Dalsgaard}, {Fischer}, {Hekker}, {Kjeldsen}, {Mathur}, {Mosser},
  {Pinsonneault}, {Valenti}, {Christiansen}, {Kinemuchi}, \&
  {Mullally}}]{Deheuvels2012}
{Deheuvels}, S., {Garc{\'\i}a}, R.~A., {Chaplin}, W.~J., {et~al.} 2012, \apj,
  756, 19

\bibitem[{{Di Mauro} {et~al.}(2016){Di Mauro}, {Ventura}, {Cardini}, {Stello},
  {Christensen-Dalsgaard}, {Dziembowski}, {Patern{\`o}}, {Beck}, {Bloemen},
  {Davies}, {De Smedt}, {Elsworth}, {Garc{\'\i}a}, {Hekker}, {Mosser}, \&
  {Tkachenko}}]{DiMauro2016}
{Di Mauro}, M.~P., {Ventura}, R., {Cardini}, D., {et~al.} 2016, \apj, 817, 65

\bibitem[{{Di Mauro} {et~al.}(2018){Di Mauro}, {Ventura}, {Corsaro}, \&
  {Lustosa De Moura}}]{DiMauro2018}
{Di Mauro}, M.~P., {Ventura}, R., {Corsaro}, E., \& {Lustosa De Moura}, B.
  2018, \apj, 862, 9

\bibitem[{{Fellay} {et~al.}(2021){Fellay}, {Buldgen}, {Eggenberger}, {Khan},
  {Salmon}, {Miglio}, \& {Montalb{\'a}n}}]{Fellay2021}
{Fellay}, L., {Buldgen}, G., {Eggenberger}, P., {et~al.} 2021, \aap, 654, A133

\bibitem[{{Foreman-Mackey} {et~al.}(2013){Foreman-Mackey}, {Hogg}, {Lang}, \&
  {Goodman}}]{Foreman2013}
{Foreman-Mackey}, D., {Hogg}, D.~W., {Lang}, D., \& {Goodman}, J. 2013, \pasp,
  125, 306

\bibitem[{{Fuller} {et~al.}(2019){Fuller}, {Piro}, \& {Jermyn}}]{Fuller2019}
{Fuller}, J., {Piro}, A.~L., \& {Jermyn}, A.~S. 2019, \mnras, 485, 3661

\bibitem[{{Garc{\'\i}a} {et~al.}(2014){Garc{\'\i}a}, {Ceillier}, {Salabert},
  {Mathur}, {van Saders}, {Pinsonneault}, {Ballot}, {Beck}, {Bloemen},
  {Campante}, {Davies}, {do Nascimento}, {Mathis}, {Metcalfe}, {Nielsen},
  {Su{\'a}rez}, {Chaplin}, {Jim{\'e}nez}, \& {Karoff}}]{Garcia2014}
{Garc{\'\i}a}, R.~A., {Ceillier}, T., {Salabert}, D., {et~al.} 2014, \aap, 572,
  A34

\bibitem[{{Gehan} {et~al.}(2018){Gehan}, {Mosser}, {Michel}, {Samadi}, \&
  {Kallinger}}]{Gehan2018}
{Gehan}, C., {Mosser}, B., {Michel}, E., {Samadi}, R., \& {Kallinger}, T. 2018,
  \aap, 616, A24

\bibitem[{Gelman {et~al.}(2013)Gelman, Carlin, Stern, Dunson, Vehtari, \&
  Rubin}]{Gelman2013}
Gelman, A., Carlin, J., Stern, H., {et~al.} 2013, Bayesian Data Analysis, Third
  Edition, Chapman \& Hall/CRC Texts in Statistical Science (Taylor \& Francis)

\bibitem[{{Gough}(1981)}]{Gough1981}
{Gough}, D.~O. 1981, \mnras, 196, 731

\bibitem[{{Gough} \& {McIntyre}(1998)}]{Gough1998}
{Gough}, D.~O. \& {McIntyre}, M.~E. 1998, \nat, 394, 755

\bibitem[{{Goupil} {et~al.}(2024){Goupil}, {Catala}, {Samadi}, {Belkacem},
  {Ouazzani}, {Reese}, {Appourchaux}, {Mathur}, {Cabrera}, {B{\"o}rner},
  {Paproth}, {Moedas}, {Verma}, {Lebreton}, {Deal}, {Ballot}, {Chaplin},
  {Christensen-Dalsgaard}, {Cunha}, {Lanza}, {Miglio}, {Morel}, {Serenelli},
  {Mosser}, {Creevey}, {Moya}, {Garcia}, {Nielsen}, \& {Hatt}}]{Goupil2024}
{Goupil}, M.~J., {Catala}, C., {Samadi}, R., {et~al.} 2024, \aap, 683, A78

\bibitem[{{Hansen} {et~al.}(1977){Hansen}, {Cox}, \& {van Horn}}]{Hansen1977}
{Hansen}, C.~J., {Cox}, J.~P., \& {van Horn}, H.~M. 1977, \apj, 217, 151

\bibitem[{{Hatta} {et~al.}(2022){Hatta}, {Sekii}, {Benomar}, \&
  {Takata}}]{Hatta2022}
{Hatta}, Y., {Sekii}, T., {Benomar}, O., \& {Takata}, M. 2022, \apj, 927, 40

\bibitem[{{Hatta} {et~al.}(2019){Hatta}, {Sekii}, {Takata}, \&
  {Kurtz}}]{Hatta2019}
{Hatta}, Y., {Sekii}, T., {Takata}, M., \& {Kurtz}, D.~W. 2019, \apj, 871, 135

\bibitem[{{Kissin} \& {Thompson}(2015)}]{Kissin2015}
{Kissin}, Y. \& {Thompson}, C. 2015, \apj, 808, 35

\bibitem[{{Maeder}(2009)}]{Maeder2009}
{Maeder}, A. 2009, {Physics, Formation and Evolution of Rotating Stars}

\bibitem[{{Marques} {et~al.}(2013){Marques}, {Goupil}, {Lebreton}, {Talon},
  {Palacios}, {Belkacem}, {Ouazzani}, {Mosser}, {Moya}, {Morel}, {Pichon},
  {Mathis}, {Zahn}, {Turck-Chi{\`e}ze}, \& {Nghiem}}]{Marques2013}
{Marques}, J.~P., {Goupil}, M.~J., {Lebreton}, Y., {et~al.} 2013, \aap, 549,
  A74

\bibitem[{{Mosser} {et~al.}(2012){Mosser}, {Goupil}, {Belkacem}, {Marques},
  {Beck}, {Bloemen}, {De Ridder}, {Barban}, {Deheuvels}, {Elsworth}, {Hekker},
  {Kallinger}, {Ouazzani}, {Pinsonneault}, {Samadi}, {Stello}, {Garc{\'\i}a},
  {Klaus}, {Li}, {Mathur}, \& {Morris}}]{Mosser2012}
{Mosser}, B., {Goupil}, M.~J., {Belkacem}, K., {et~al.} 2012, \aap, 548, A10

\bibitem[{{Pijpers} \& {Thompson}(1994)}]{Pijpers}
{Pijpers}, F.~P. \& {Thompson}, M.~J. 1994, \aap, 281, 231

\bibitem[{{Pin{\c{c}}on} {et~al.}(2017){Pin{\c{c}}on}, {Belkacem}, {Goupil}, \&
  {Marques}}]{Pincon2017}
{Pin{\c{c}}on}, C., {Belkacem}, K., {Goupil}, M.~J., \& {Marques}, J.~P. 2017,
  \aap, 605, A31

\bibitem[{{Rabello-Soares} {et~al.}(1999){Rabello-Soares}, {Basu}, \&
  {Christensen-Dalsgaard}}]{RabelloParam}
{Rabello-Soares}, M.~C., {Basu}, S., \& {Christensen-Dalsgaard}, J. 1999,
  MNRAS, 309, 35

\bibitem[{{Rauer} {et~al.}(2014){Rauer}, {Catala}, {Aerts}, {Appourchaux},
  {Benz}, {Brandeker}, {Christensen-Dalsgaard}, {Deleuil}, {Gizon}, {Goupil},
  {G{\"u}del}, {Janot-Pacheco}, {Mas-Hesse}, {Pagano}, {Piotto}, {Pollacco},
  {Santos}, {Smith}, {Su{\'a}rez}, {Szab{\'o}}, {Udry}, {Adibekyan}, {Alibert},
  {Almenara}, {Amaro-Seoane}, {Eiff}, {Asplund}, {Antonello}, {Barnes},
  {Baudin}, {Belkacem}, {Bergemann}, {Bihain}, {Birch}, {Bonfils}, {Boisse},
  {Bonomo}, {Borsa}, {Brand{\~a}o}, {Brocato}, {Brun}, {Burleigh}, {Burston},
  {Cabrera}, {Cassisi}, {Chaplin}, {Charpinet}, {Chiappini}, {Church},
  {Csizmadia}, {Cunha}, {Damasso}, {Davies}, {Deeg}, {D{\'\i}az}, {Dreizler},
  {Dreyer}, {Eggenberger}, {Ehrenreich}, {Eigm{\"u}ller}, {Erikson}, {Farmer},
  {Feltzing}, {de Oliveira Fialho}, {Figueira}, {Forveille}, {Fridlund},
  {Garc{\'\i}a}, {Giommi}, {Giuffrida}, {Godolt}, {Gomes da Silva}, {Granzer},
  {Grenfell}, {Grotsch-Noels}, {G{\"u}nther}, {Haswell}, {Hatzes},
  {H{\'e}brard}, {Hekker}, {Helled}, {Heng}, {Jenkins}, {Johansen},
  {Khodachenko}, {Kislyakova}, {Kley}, {Kolb}, {Krivova}, {Kupka}, {Lammer},
  {Lanza}, {Lebreton}, {Magrin}, {Marcos-Arenal}, {Marrese}, {Marques},
  {Martins}, {Mathis}, {Mathur}, {Messina}, {Miglio}, {Montalban}, {Montalto},
  {Monteiro}, {Moradi}, {Moravveji}, {Mordasini}, {Morel}, {Mortier},
  {Nascimbeni}, {Nelson}, {Nielsen}, {Noack}, {Norton}, {Ofir}, {Oshagh},
  {Ouazzani}, {P{\'a}pics}, {Parro}, {Petit}, {Plez}, {Poretti}, {Quirrenbach},
  {Ragazzoni}, {Raimondo}, {Rainer}, {Reese}, {Redmer}, {Reffert},
  {Rojas-Ayala}, {Roxburgh}, {Salmon}, {Santerne}, {Schneider}, {Schou},
  {Schuh}, {Schunker}, {Silva-Valio}, {Silvotti}, {Skillen}, {Snellen}, {Sohl},
  {Sousa}, {Sozzetti}, {Stello}, {Strassmeier}, {{\v{S}}vanda}, {Szab{\'o}},
  {Tkachenko}, {Valencia}, {Van Grootel}, {Vauclair}, {Ventura}, {Wagner},
  {Walton}, {Weingrill}, {Werner}, {Wheatley}, \& {Zwintz}}]{Rauer2014}
{Rauer}, H., {Catala}, C., {Aerts}, C., {et~al.} 2014, Experimental Astronomy,
  38, 249

\bibitem[{{Reese}(2018)}]{Reese2018}
{Reese}, D.~R. 2018, in Astrophysics and Space Science Proceedings, Vol.~49,
  Asteroseismology and Exoplanets: Listening to the Stars and Searching for New
  Worlds, ed. T.~L. {Campante}, N.~C. {Santos}, \& M.~J.~P.~F.~G. {Monteiro},
  75

\bibitem[{{Ricker} {et~al.}(2015){Ricker}, {Winn}, {Vanderspek}, {Latham},
  {Bakos}, {Bean}, {Berta-Thompson}, {Brown}, {Buchhave}, {Butler}, {Butler},
  {Chaplin}, {Charbonneau}, {Christensen-Dalsgaard}, {Clampin}, {Deming},
  {Doty}, {De Lee}, {Dressing}, {Dunham}, {Endl}, {Fressin}, {Ge}, {Henning},
  {Holman}, {Howard}, {Ida}, {Jenkins}, {Jernigan}, {Johnson}, {Kaltenegger},
  {Kawai}, {Kjeldsen}, {Laughlin}, {Levine}, {Lin}, {Lissauer}, {MacQueen},
  {Marcy}, {McCullough}, {Morton}, {Narita}, {Paegert}, {Palle}, {Pepe},
  {Pepper}, {Quirrenbach}, {Rinehart}, {Sasselov}, {Sato}, {Seager},
  {Sozzetti}, {Stassun}, {Sullivan}, {Szentgyorgyi}, {Torres}, {Udry}, \&
  {Villasenor}}]{Ricker2015}
{Ricker}, G.~R., {Winn}, J.~N., {Vanderspek}, R., {et~al.} 2015, Journal of
  Astronomical Telescopes, Instruments, and Systems, 1, 014003

\bibitem[{{Sekii}(1997)}]{Sekii1997}
{Sekii}, T. 1997, in Sounding Solar and Stellar Interiors, ed. J.~{Provost} \&
  F.-X. {Schmider}, Vol. 181, ISBN0792348389

\bibitem[{{Spruit}(2002)}]{spr02}
{Spruit}, H.~C. 2002, \aap, 381, 923

\bibitem[{{Takahashi} \& {Langer}(2021)}]{Takahashi2021}
{Takahashi}, K. \& {Langer}, N. 2021, \aap, 646, A19

\bibitem[{{Wilson} {et~al.}(2023){Wilson}, {Casey}, {Mandel}, {Ball},
  {Bellinger}, \& {Davies}}]{Wilson2023}
{Wilson}, T.~A., {Casey}, A.~R., {Mandel}, I., {et~al.} 2023, \mnras, 521, 4122

\end{thebibliography}
\appendix

\section{Posterior distributions for the MCMC runs}\label{sec:Appendix}

In this section, we provide the full posterior distributions of the parameters for the Gaussian profiles for Stars A,D,F, Step profiles for Stars B and E and Extended Power Law parametric profiles for Star C. 

\begin{figure*}
	\centering
		\includegraphics[width=10cm]{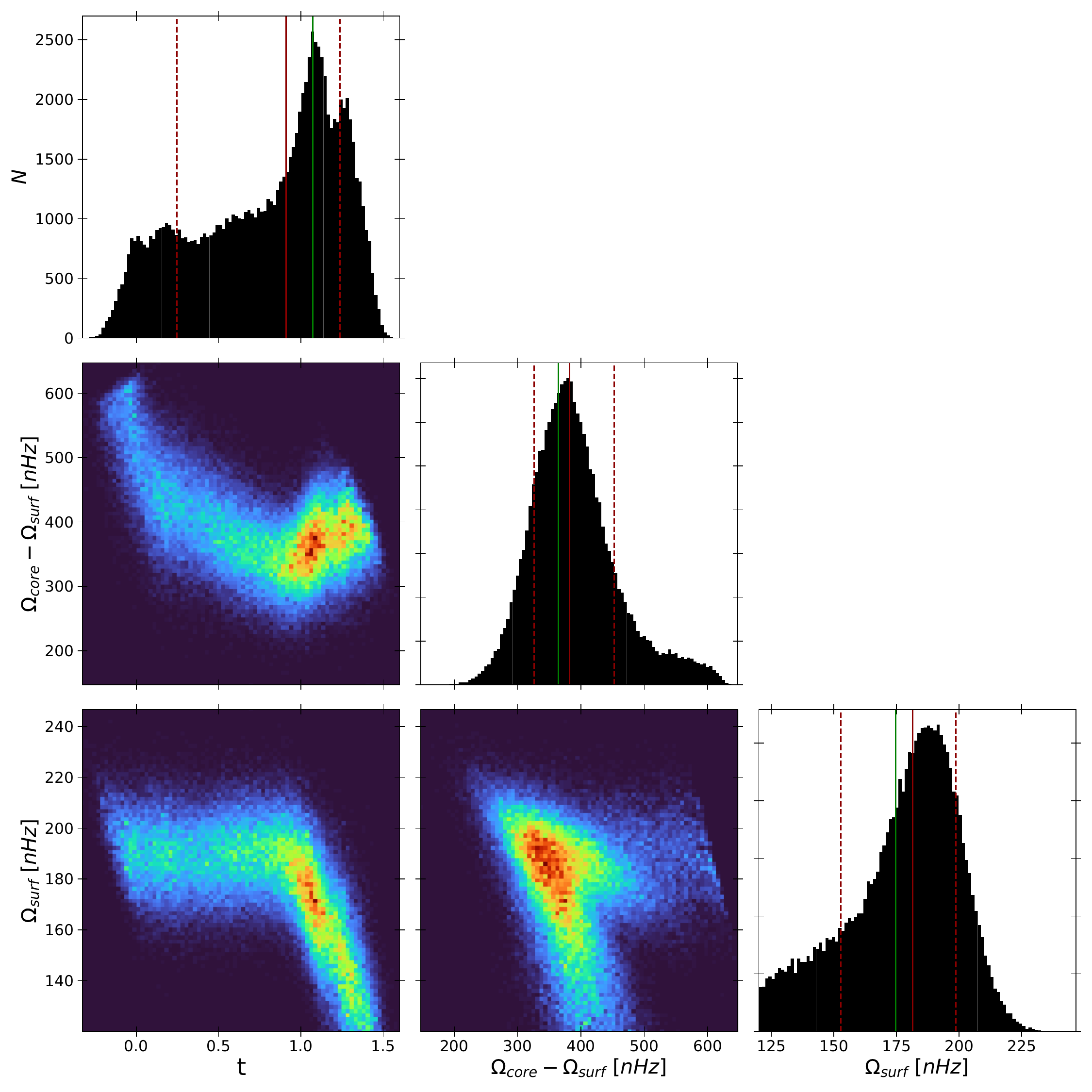}
	\caption{Posterior distributions for the MCMC inversion of Star A using the Gaussian rotation profile.}
		\label{Fig:DistribA}
\end{figure*} 
\begin{figure*}
	\centering
		\includegraphics[width=10cm]{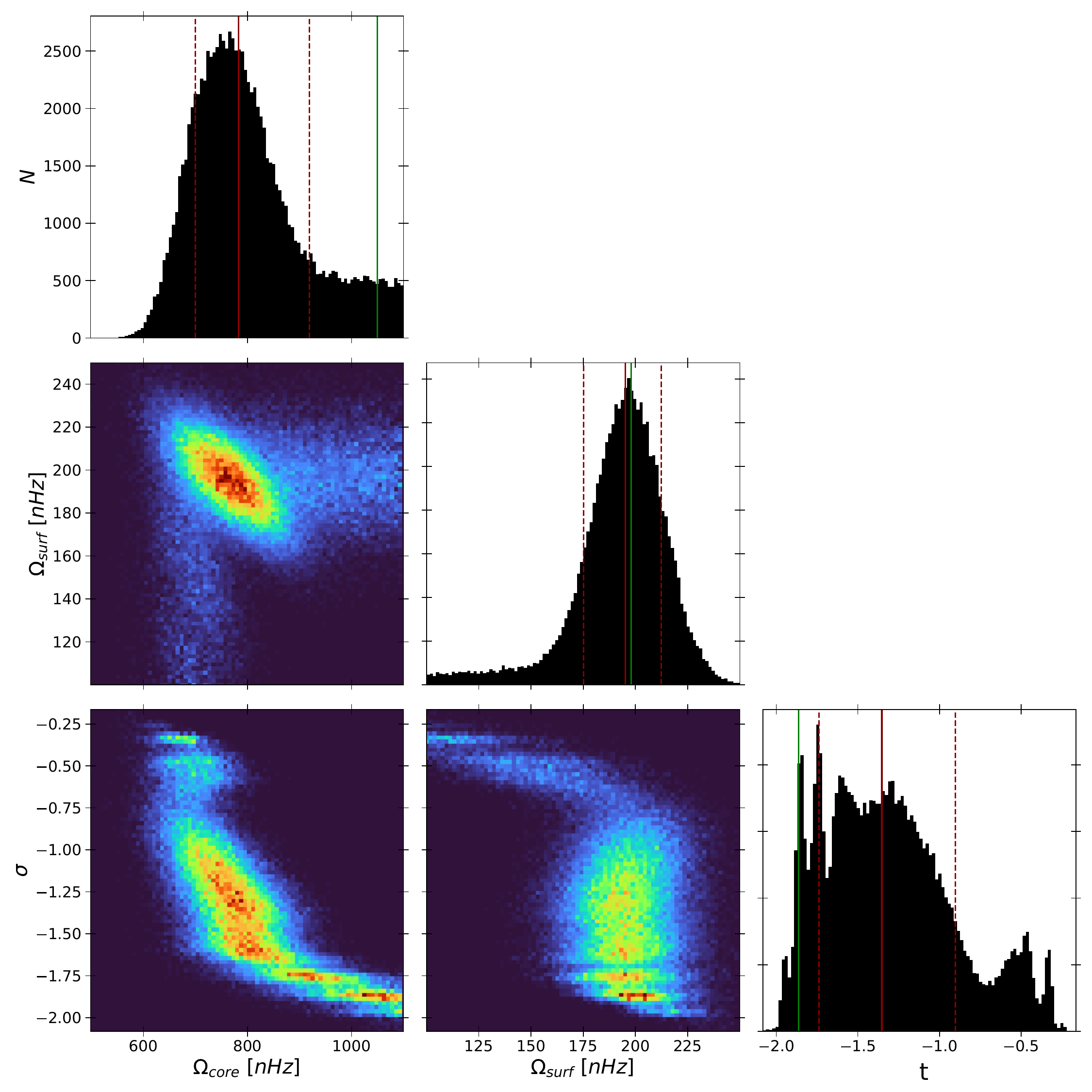}
	\caption{Posterior distributions for the MCMC inversion of Star B using the step rotation profile.}
		\label{Fig:DistribB}
\end{figure*} 
\begin{figure*}
	\centering
		\includegraphics[width=10cm]{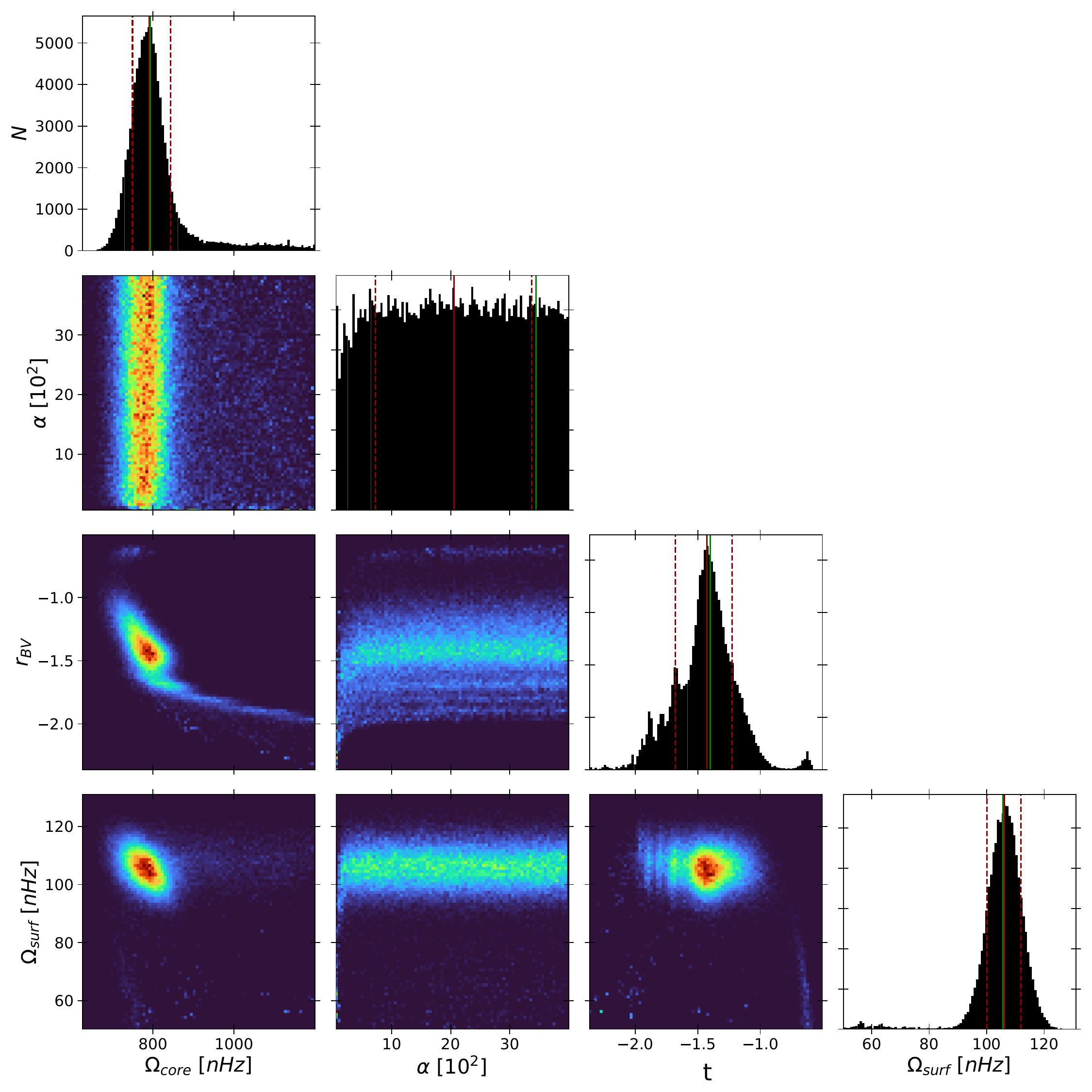}
	\caption{Posterior distributions for the MCMC inversion of Star C using the extended power law rotation profile.}
		\label{Fig:DistribC}
\end{figure*} 
\begin{figure*}
	\centering
		\includegraphics[width=10cm]{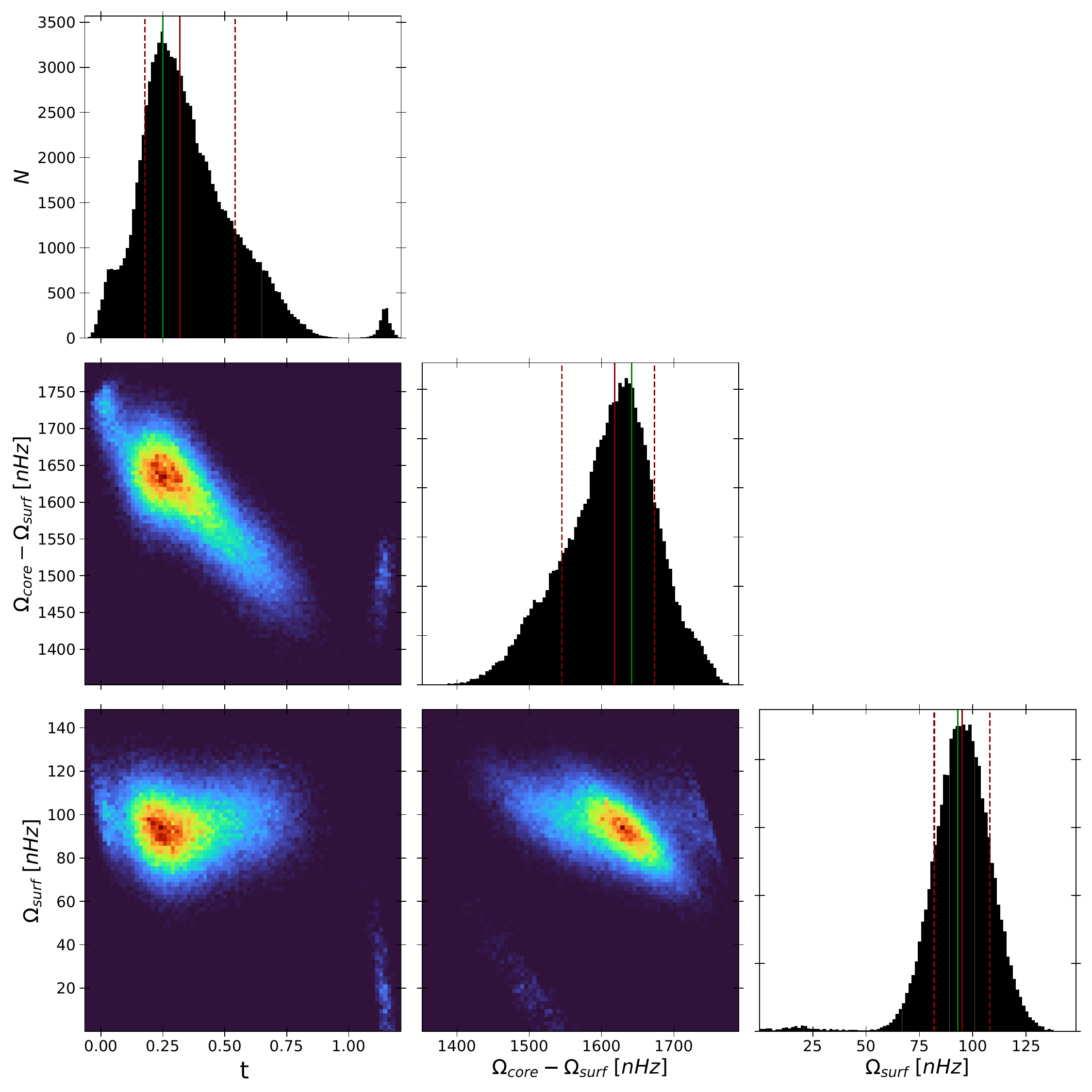}
	\caption{Posterior distributions for the MCMC inversion of Star D using the Gaussian rotation profile.}
		\label{Fig:DistribD}
\end{figure*} 
\begin{figure*}
	\centering
		\includegraphics[width=10cm]{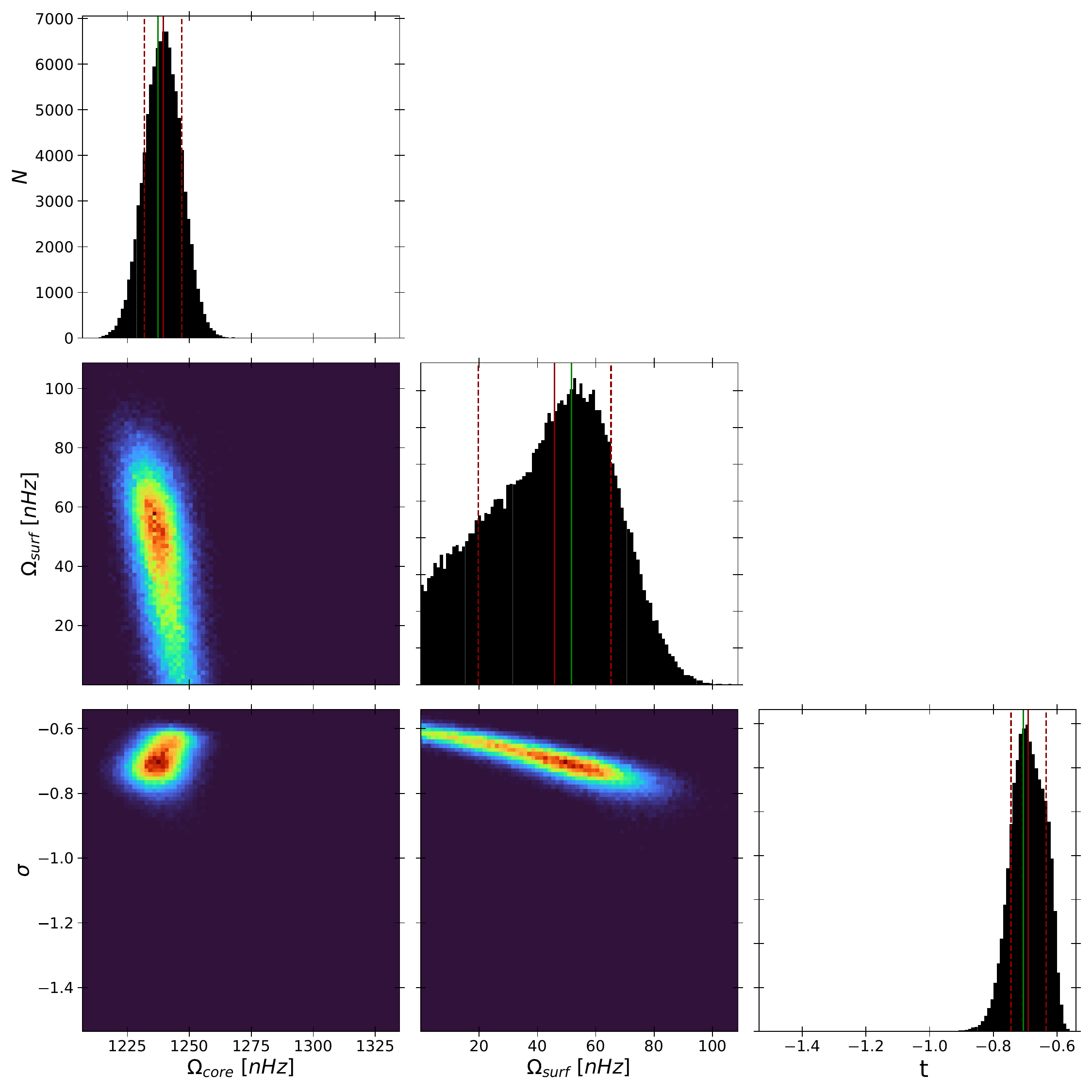}
	\caption{Posterior distributions for the MCMC inversion of Star E using the step rotation profile.}
		\label{Fig:DistribE}
\end{figure*} 
\begin{figure*}
	\centering
		\includegraphics[width=10cm]{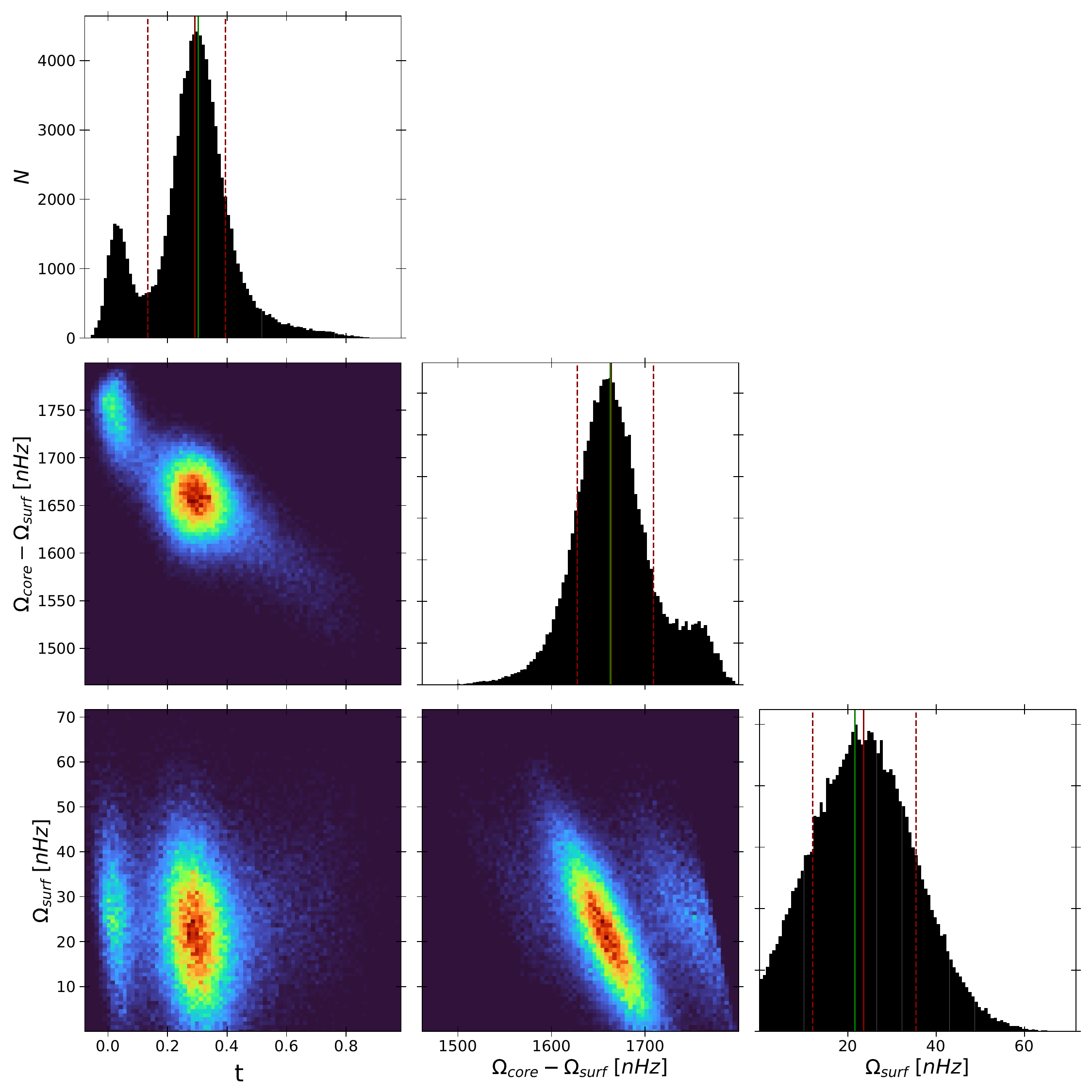}
	\caption{Posterior distributions for the MCMC inversion of Star F using the Gaussian rotation profile.}
		\label{Fig:DistribF}
\end{figure*} 
\end{document}